\documentclass[reprint,superscriptaddress,amsmath,amssymb,aps,pra,floatfix]{revtex4-1}

\usepackage{amsfonts}
\usepackage{amsmath}
\usepackage{amsthm}
\usepackage{mathrsfs}
\usepackage{amscd}
\usepackage{amssymb}
\usepackage{subfigure}
\usepackage{amsxtra}
\usepackage{dcolumn}% Align table columns on decimal point
\usepackage{bm}           % bold math
\usepackage{bbm}
\usepackage{graphicx}
\usepackage{epstopdf}

\usepackage[colorlinks]{hyperref}

\newcommand{\la}{\langle}
\newcommand{\ra}{\rangle}
\newcommand{\bra}[1]{\ensuremath{\langle#1|}}
\newcommand{\ket}[1]{\ensuremath{\left|#1\right\rangle}}

\newcommand{\ketbra}[2]{\ensuremath{\left| #1 \right\rangle \left\langle #2 \right|}}

\newcommand{\cL}{\mathcal{L}}
\newcommand{\cD}{\mathcal{D}}

\newcommand{\dg}{\dagger}
\newcommand{\da}{\dagger}

\newcommand{\Op}[1]{\hat{#1}}

\newcommand{\oL}{\Op{L}}
\newcommand{\oH}{\Op{H}}
\newcommand{\oA}{\Op{A}}
\newcommand{\oB}{\Op{B}}
\newcommand{\oU}{\Op{U}}

\newcommand{\oS}{\Op{S}}
\newcommand{\oJ}{\Op{J}}
\newcommand{\oj}{\Op{j}}

\newcommand{\ovJ}{\Op{\mathbf{J}}}
\newcommand{\ovj}{\Op{\mathbf{j}}}
\newcommand{\vxi}{\boldsymbol{\xi}}
\newcommand{\id}{\ensuremath{\mathbbm 1}}
\newcommand{\tr}{\ensuremath{{\rm tr}}}
\renewcommand{\Im}{\ensuremath{{\rm Im}}}
\newcommand{\dd}{\mathrm{d}}

\newcommand{\half}{\tfrac{1}{2}}

\begin{document}

\preprint{APS/123-QED}

\title{Nonequilibrium Dynamics with Finite-Time Repeated Interactions}
\author{Stella Seah}
\affiliation{Department of Physics, National University of Singapore, 2 Science Drive 3, Singapore 117542, Singapore}

\author{Stefan Nimmrichter}
\affiliation{Centre for Quantum Technologies, National University of Singapore, 3 Science Drive 2, Singapore 117543, Singapore}

\author{Valerio Scarani}
\affiliation{Department of Physics, National University of Singapore, 2 Science Drive 3, Singapore 117542, Singapore}
\affiliation{Centre for Quantum Technologies, National University of Singapore, 3 Science Drive 2, Singapore 117543, Singapore}

\date{\today}% It is always \today, today,
             %  but any date may be explicitly specified
             
\begin{abstract}
We study quantum dynamics in the framework of repeated interactions between a system and a stream of identical probes. We present a coarse-grained master equation that captures the system's dynamics in the natural regime where interactions with different probes do not overlap, but is otherwise valid for arbitrary values of the interaction strength and mean interaction time. We then apply it to some specific examples. For probes prepared in Gibbs states, such channels have been used to describe thermalisation: while this is the case for many choices of parameters, for others one finds out-of-equilibrium states including inverted Gibbs and maximally mixed states. Gapless probes can be interpreted as performing an indirect measurement, and we study the energy transfer associated with this measurement. 

\end{abstract}
\maketitle

\section{Introduction}\label{sec:intro}
In the study of open quantum systems, various schemes were conceived and employed to describe dynamics of a physical system under the influence of uncontrollable degrees of freedom from its environment. For instance, this influence may be broken down into individual events, where the system is assumed to interact with an environmental probe (or ancilla). Such repeated interaction models have been studied previously in the context of heat dissipation and thermalization \cite{scarani2002,landi2014,lorenzo2015,grimmer2016,Baumer2017,Strasberg2017,Hanson2018,Grimmer2018,Grimmer2018b,deChiara2018} as well as applications in thermal machines \cite{barra2015,hardal2015superradiant,manzano2017,Strasberg2017,Barra2018,Pezzutto2018}. However, an effective master equation description has been proposed only in the limit of short (and strong) system-probe interactions 
\cite{landi2014,lorenzo2015,barra2015,grimmer2016,Strasberg2017,Grimmer2018,Grimmer2018b,deChiara2018}.

Here we formulate a scheme that extends this framework to arbitrary interaction strengths and times (Sec.~\ref{sec:model}), resulting in features missed in previous short-time treatments. After some general observations (Sec.~\ref{sec:general}), we illustrate this by first considering repeated interactions with thermal probes in Sec.~\ref{sec:1SpinThermal}. We show how systems can achieve tunable out-of-equilibrium steady states, including population inversion, depending on the interaction time. In fact, the system will almost never equilibrate to the temperature of the environment unless the interactions are strictly energy-preserving. Specifically, we shall see in Sec.~\ref{sec:composite} that a composite system, regardless of its internal coupling, can never thermalize as a whole and would at best equilibrate locally should the probes interact locally with one subsystem. As a second example, we look at repeated interactions in the form of an indirect measurement process in Sec.~\ref{sec:meas}. We shall see that only gapless probes can realize ideal von Neumann measurements, whereas realistic probe interactions result in a more complex system evolution. Our model also sheds light on the origin of the apparent heating effect, which is simply the external work associated to switching on and off the interaction between the system and the probes.

\section{Repeated interaction model}\label{sec:model}

We consider a system with Hamiltonian $\oH_s$ that interacts with a sequence of non-interacting probes with Hamiltonians $\oH_p$ in discrete, non-overlapping events. Each interaction is given by the unitary $\oU (\tau) = \exp [- i ( \oH_0 + \oH_{\rm int} )\tau/\hbar]$ with $\oH_0 = \oH_s + \oH_p$. It describes a coupling Hamiltonian $\oH_{\rm int}$ with characteristic coupling strength $g$ that is switched on for a finite duration $\tau$. 
If the interaction is repeated at \emph{regular} time steps $\Delta t$ with identical probe units in states $\eta$, the system state $\rho$ evolves through applications of the map \cite{barra2015,Strasberg2017}
\begin{equation}
    \Phi(\rho) = \tr_p \left[\oU_0\left(\Delta t-\tau\right) \oU\left(\tau\right)(\rho\otimes\eta) \oU^\dg\left(\tau\right) \oU_0^\dg\left(\Delta t-\tau\right)\right].
\end{equation}
This process can be viewed as an incoherent energy exchange channel, but it can also describe external driving through periodic (and possibly resonant) control pulses, the period set by the waiting time $\Delta t$ between successive events.

\begin{figure}
\centerline{\includegraphics[width=\columnwidth]{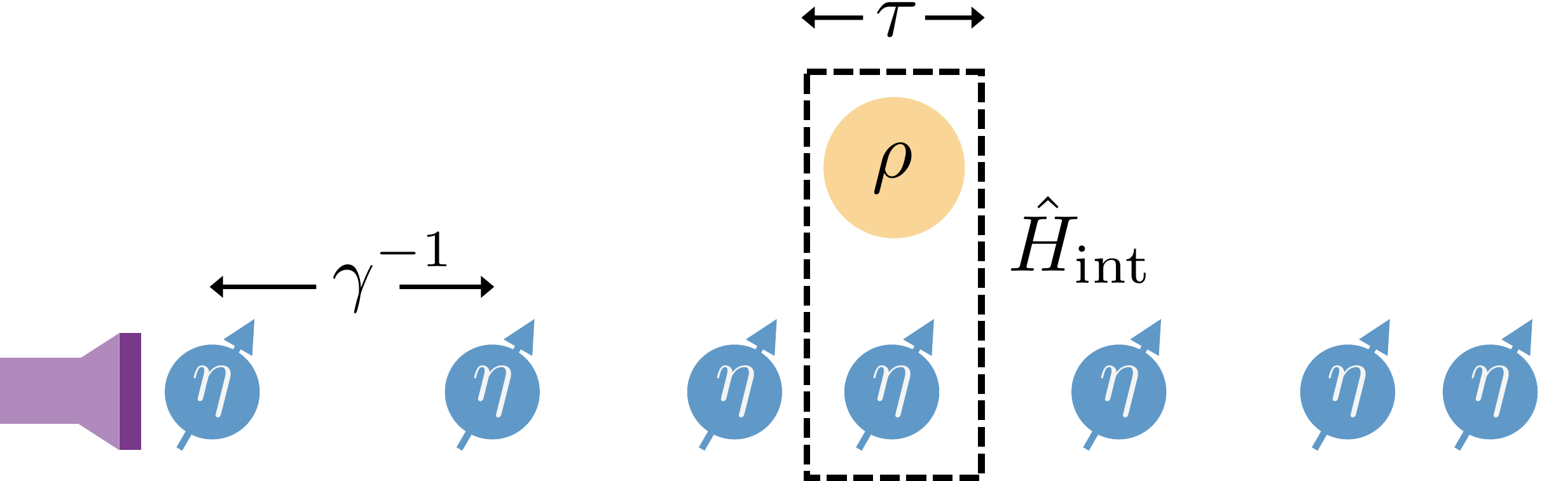}}
\caption{\label{fig:sketch}Sketch of a random repeated interaction process. A system state $\rho$ interacts sequentially at an average rate $\gamma$ with individual probes in state $\eta$. Each interaction is mediated by a Hamiltonian $\oH_{\rm int}$ switched on for a duration $\tau$. }
\end{figure}

In situations where the exact timing of an interaction event is irrelevant or unknown, we could assume a Poisson process that occurs at an average rate $\gamma$, as sketched in Fig.~\ref{fig:sketch}. This would allow us to derive a master equation that describes the system evolution based on the stochastic jump processes induced by the repeated interactions. The probes may represent uncontrollable degrees of freedom of a thermal environment, in which case $\eta = \exp[-\beta \oH_p]/Z_p$. 
The waiting time $\Delta t$ is then a random variable following an exponential distribution $p(\Delta t) = \gamma e^{-\gamma \Delta t}$, and we demand that $\gamma\tau\ll 1$ in order to stay within the Markovian framework of non-overlapping interactions.
While this implies that the system evolves freely for most of the time and gets interrupted only by occasional events, the latter may well describe \emph{strong} system-probe interactions. This contrasts the weak coupling assumption necessary for obtaining standard Born-Markov master equations.

To arrive at a consistent description of the system evolution across arbitrary interaction strengths $g$ and times $\tau$, we employ a formalism inspired by scattering theory and collisional decoherence \cite{Gallis1990,Diosi1995,Hornberger2003c,Hornberger2006,Hornberger2007,Vacchini2009} that accurately describes each jump by singling out the net effect of the interaction Hamiltonian from the free evolution,
\begin{equation} 
\oS = \oU_{0} \left(-\frac{\tau}{2}\right)\oU (\tau) \oU_{0} \left(-\frac{\tau}{2}\right).
\label{eq:Smatrix}
\end{equation} 
The symmetric form of the product implies that $\oS^\da$ describes the time-reversed event. 

The average change in system state, coarse-grained over individual interaction events is given by 
\begin{equation}
\dd{\rho} = \left\{-\frac{i}{\hbar} [\oH_s,\rho] \dd t \right\} (1-\gamma\, \dd t) + \gamma\, \dd t \left[ \tr_p \{ \oS \rho \otimes \eta \oS^\da \} - \rho \right],
\end{equation}
provided that $\gamma\tau \ll 1$. Keeping terms of $\dd t$ up to first order, we arrive at a master equation 
\begin{equation}
\dot{\rho} = -\frac{i}{\hbar} [\oH_s,\rho] + \gamma \left[ \tr_p \{ \oS \rho \otimes \eta \oS^\da \} - \rho \right], \label{eq:SmatrixME}
\end{equation}
Unlike previous studies, where master equations are derived expanding $\oU(\tau)$ in the limit of short ($g\tau\ll 1$) and possibly strong ($g\gg \omega_{s,p}$) interactions \cite{landi2014,lorenzo2015,barra2015,Strasberg2017,grimmer2016,Grimmer2018,Grimmer2018b,deChiara2018}, our approach captures the dynamics of the process without imposing any parameter constraints other than $\gamma\tau\ll 1$ (see App.~\ref{app:simulation}).
Notice that \eqref{eq:SmatrixME} is of Lindblad form: indeed, with $\eta = \sum_k \eta_k |k\ra\la k|$ and $\oL_{k\ell} = \la \ell | \oS | k\ra$, it reads
\begin{equation}
\dot{\rho} = -\frac{i}{\hbar} [\oH_s,\rho] + \gamma \sum_{k,\ell} \eta_k \left[ \oL_{k \ell} \rho \oL_{k \ell}^\da - \tfrac{1}{2} \{ \oL_{k \ell}^\da \oL_{k \ell},\rho \} \right]. \label{eq:SmatrixME_Lind}
\end{equation}

This framework of repeated interactions could be extended beyond the assumption of identical events by introducing a multidimensional random variable $\vxi$, which accounts for the fluctuations in $\{\oH_p (\vxi),\oH_{\rm int} (\vxi),\eta(\vxi),\tau(\vxi) \}$ arising from inhomogeneity of the probes or of the interaction events. The dissipative part of the master equation \eqref{eq:SmatrixME} would then be replaced by the ensemble-averaged expression
\begin{equation}
\cL \rho = \gamma \Big[ \sum_{\vxi} p(\vxi) \tr_p \left\{ \oS (\vxi) \rho \otimes \eta (\vxi) \oS^\da (\vxi) \right\} - \rho \Big], \label{eq:SmatrixMEfluct}
\end{equation} where $p(\vxi)$ is probability distribution of $\vxi$.
In case it is only the probe states $\eta(\xi)$ that fluctuate, we can introduce the ensemble-averaged state $\eta = \sum_{\xi} p(\xi) \eta (\xi)$ to retrieve the simpler form \eqref{eq:SmatrixME}.

\section{General considerations}\label{sec:general}

Before studying specific models of repeated interactions, we present two general considerations: a connection with the resource theory approach to thermodynamics, and some features of the dynamics in the limit of short interaction time $\tau$.

\subsection{Repeated interactions and models of thermalisation} \label{subsec:thermo}

A frequently studied family of interactions are the so-called energy-preserving interactions, for which there is no net work cost in coupling the system to the probe. Mathematically, this translates as $[\oU (\tau), \oH_0]=0$, that is $[\oH_{\rm int}, \oH_0]=0$ if we want the condition to be valid for all $\tau$. In other words, energy-preserving interactions can only mediate an exchange of excitations between the system and the probe. For these interactions, \eqref{eq:Smatrix} reduces to the eikonal expression
\begin{equation} \label{eq:SmatrixTO}
\oS_{\rm eik} = \exp \left[ -\frac{i \tau}{\hbar} \oH_{\rm int} \right].  
\end{equation} Repeated-interaction channels with energy-preserving interactions and with the probe in the thermal state are a realisation of \emph{``thermal operations''} i.e.~free operations in the framework of thermal resource theory \cite{Brandao2013,Brandao2015,Gour2015,goold2016review,Lostaglio2018}. These channels bring the system state closer to (or at least not further away from) its Gibbs state. Indeed, the commutation requirement implies that $\rho = \exp [-\beta \oH_s]/Z_s$ is stationary under \eqref{eq:SmatrixME}. 

Conversely, for interactions that are not energy-preserving, repeated interactions will typically lead to equilibration to a non-thermal steady state. In such cases, additional work is required to mediate a single interaction event between the system and probe \cite{Strasberg2017}.
The average work power associated to the coarse-grained time evolution is then given by the rate of energy change 
\begin{eqnarray}\label{eq:workPower}
\dot{W} &=& \gamma \tr \left\{ \oU_0 \left( \frac{\tau}{2} \right) \rho \otimes \eta  \oU_0 \left( - \frac{\tau}{2} \right) \left[ \oS^\da \oH_0 \oS - \oH_0 \right] \right\}. 
\end{eqnarray}

\subsection{Short-time limit}\label{sec:shorttime}

Previous works \cite{landi2014,lorenzo2015,barra2015,Strasberg2017,grimmer2016,Grimmer2018,Grimmer2018b,deChiara2018} on repeated interactions mainly focus on the short-time limit, where $\oS_{\rm{eik}}$ is used to capture the dynamics between the system and the probes. Here, we see that using the proposed scattering operator $\oS$, we are able to retrieve the same results regardless of whether the interaction is energy-preserving, as follows from a Baker-Campbell-Hausdorff expansion of \eqref{eq:Smatrix} \cite{Yoshida1990}, 
\begin{eqnarray}
\ln \oS &=& -\frac{i \tau}{\hbar} \left( \oH_{\rm int} + \frac{\tau^2}{12\hbar^2} [\oH_{\rm int}, [\oH_{\rm int},\oH_0]] \right.  \nonumber \\
&& \left. - \frac{\tau^2}{24\hbar^2} [\oH_0, [\oH_0,\oH_{\rm int}]] + \ldots \right)
\end{eqnarray}
The scattering operator \eqref{eq:SmatrixTO} can then be Taylor-expanded to recover the known short-time master equation .
However, the condition $\omega_s \tau \ll 1$ explicitly breaks the validity of the rotating wave approximation (RWA), as it makes a difference whether the full interaction Hamiltonian or only its resonant terms are plugged into the short-time operator \eqref{eq:SmatrixTO}. In particular, caution must be exercised in such cases when the system-probe interaction $\oH_{\rm int}$ does not preserve energy in the first place, but it could be reduced to its energy preserving resonant terms by exploiting the RWA. This may lead to wrong conclusions for the effect of short-time interactions.

Consider for instance the short-time behavior of product interactions, $\oH_{\rm int} = \hbar g \oA \otimes \oB$, where $\oA$ and $\oB$ are Hermitian operators. Expanding the bath operator into its eigenbasis, $\oB = \sum_k b_k |k\ra\la k|$, we find that the single events are described by a mixture of unitary transformations of the system state, 
\begin{equation}
\tr_p \{ \oS_{\rm eik} \rho \otimes \eta \oS_{\rm eik}^\da \} = \sum_k \la k |\eta |k\ra e^{-i g b_k \tau \oA} \rho e^{i g b_k \tau \oA}.
\end{equation}
The von Neumann entropy, a concave function of the system state, can only increase under such an operation: 
a purity-decreasing unital map \cite{Lidar2006,Grimmer2017a}.
Hence, repeated product interactions could not describe thermalization to a finite temperature, as such a process would be able to decrease the entropy of states that are initially hotter. We are left instead with combinations of dephasing and heating towards the maximally mixed state; energy preserving product interactions would result in pure dephasing. 

Conversely, thermalizing master equations can be consistently obtained from a product interaction by performing the secular approximation in the standard Born-Markov approach \cite{Breuer2002}. As we shall demonstrate, repeated interactions will act like conventional heat reservoirs and lead to the correct thermal state only in exceptional cases of resonant energy exchanges. 
In other cases, they serve as ergotropy reservoirs that lead to out-of-equilibrium states including population inversion.
For purely phase-modulating couplings, $[\oH_{\rm int}, \oH_s]=0$, the Born-Markov setting  and repeated interactions both result in a dephasing master equation of the same form \cite{Doll2008}.

\section{Spin equilibration with thermal probes}\label{sec:1SpinThermal}

In this section we consider a single spin $\ovJ$ with transition frequency $\omega_s$, interacting with identical probe spins $\ovj$ with transition frequency $\omega_p$. The probes are initialised as thermal states of an environment of inverse temperature $\beta$. We study the family of interaction Hamiltonians
\begin{eqnarray}
H&=&\oH_s+\oH_p+\oH_{\rm int}\nonumber\\
&=& \hbar \omega_s \oJ_z+\hbar \omega_p \oj_z+\hbar \sum_{k=x,y,z} g_k \oJ_k \otimes \oj_k\,.
\end{eqnarray} This family includes the dephasing channel ($g_{x,y}=0$), the pure exchange of excitations ($g_x=g_y$ and $g_z=0$), as well as the dipole-dipole coupling $\oH_{\rm int} \propto \ovJ \cdot \ovj$ when $g_{x,y,z} = g$. For qubits, the general master equation is presented in App.~\ref{app:XXME}. A higher system spin under repeated exchange interactions with qubits is considered in App.~\ref{app:spinME}. In the following, we illustrate the main features of equilibration under repeated linear interactions by means of the instructive and often employed case of a qubit interacting with resonant qubit probes. Higher system spins, non-resonant probes, and a composite two-spin system will be considered later.

\subsection{Qubit system and resonant probe qubits}

We now set $J=j=1/2$ and $\omega_s=\omega_p$; we also omit the pure dephasing contribution by setting $g_z=0$. The master equation \eqref{eq:SmatrixME} becomes (see App.~\ref{app:XXME}),
\begin{eqnarray}
\dot{\rho} &=& -i (\omega_s + \delta \omega) [\oJ_z,\rho] \quad  +\sum_{k=x,z,\pm} \!\! \Gamma_k \cD[\oJ_k]\rho .
\end{eqnarray}
It consists of a coherent shift $\delta \omega$ in the system frequency, a dephasing channel at the rate $\Gamma_z$, a bit flip channel at $\Gamma_x$, and incoherent energy exchange channels with the (de-)excitation rates $\Gamma_\pm$, 
\begin{eqnarray}
\delta \omega &=& \gamma \Im \{C\} \cos \frac{G_+ \tau}{2}, \\
\Gamma_z &=& \gamma \left| C -  \cos \frac{G_+ \tau}{2} \right|^2, \qquad \Gamma_x = 4\gamma K \sin \frac{G_+ \tau}{2}, \nonumber \\
\Gamma_\pm &=& \gamma \frac{e^{\pm \beta \omega_s/2}}{Z_p} \left[ K - \sin \frac{G_+ \tau}{2} \right] \left[ K - e^{\mp \beta \omega_s} \sin \frac{G_+ \tau}{2} \right]. \nonumber
\end{eqnarray}
Here we have abbreviated $G_{\pm} = (g_x \pm g_y)/2$ and we have denoted
\begin{eqnarray}
C &=& \frac{\exp \left[-i\left(2\omega_s+\sqrt{G_-^2 + 4\omega_s^2} \right)\tau/2\right] }{G_-^2 + \left(2\omega_s+\sqrt{G_-^2 + 4\omega_s^2}\right)^2} \\
&&\times \left[ e^{i \sqrt{G_-^2 + 4\omega_s^2}\tau} \left(2\omega_s+\sqrt{G_-^2 + 4\omega_s^2}\right)^2 + G_-^2 \right], \nonumber \\
K &=& \frac{2G_- \left(2\omega_s +\sqrt{G_-^2 + 4\omega_s^2}\right) }{G_-^2 + \left(2\omega_s+\sqrt{G_-^2 + 4\omega_s^2}\right)^2}\sin\frac{\sqrt{G_-^2 + 4\omega_s^2}}{2}\tau . \nonumber
\end{eqnarray}
Notice that only the exchange rates $\Gamma_\pm$ depend on the probe temperature here. $\Gamma_{z,x}$ and $\delta \omega$ would also exhibit such a dependence off-resonance or for $g_z \neq 0$.

The system qubit equilibrates to a steady state that is diagonal in the eigenbasis of $\sigma_z$, with the ratio of the probabilities of excited and ground states
\begin{equation} \label{eq:qubitRatioGen}
\chi =  \frac{\bra{\half}\rho\ket{\half}}{\bra{-\half}\rho\ket{-\half}}= \frac{e^{\beta \omega_s} K^2+ \sin^2(G_+\tau/2)}{K^2+ e^{\beta \omega_s} \sin^2 (G_+ \tau/2) }.
\end{equation} 
For interaction times $\sqrt{G_-^2 + 4\omega_s^2}\tau = 2m\pi$, $m\in \mathbb{Z}^+$, we have $K=0$ and therefore $\chi=\exp (-\beta \omega_s)$: the system converges to the Gibbs state at the temperature of the probe.
The condition $\chi=\exp (\beta \omega_s)$ means that the system converges to the inverted Gibbs state (``negative temperature''). This happens when $(g_x+g_y)\tau$ is a multiple of $4\pi$.
%The condition $\chi=\exp (\beta \omega_s)$ means that the system converges to the Gibbs state at the temperature of the probe. This happens when $(g_x+g_y)\tau$ is a multiple of $4\pi$. For interaction times $\sqrt{G_-^2 + 4\omega_s^2}\tau = 2m\pi$, $m\in \mathbb{Z}^+$, we have $K=0$ and therefore $\chi=\exp (-\beta \omega_s)$: the system converges to the inverted Gibbs state (``negative temperature'').

\begin{figure}
\centerline{\includegraphics[width=\columnwidth]{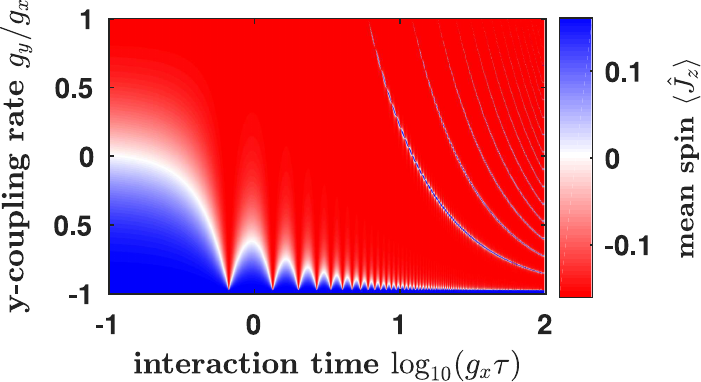}}
\caption{\label{fig:XYint} (color online) Steady state $\la \oJ_z \ra$ for $J = j = 1/2$ for different interaction times $\tau $ at a rate $\gamma =10^{-3}g_x $. Here, $\omega_p=\omega_s=4.4g_x$ and $g_z = 0$ and we consider thermal probes at $k_BT = 1.5 \hbar\omega_p$. The large red area and the smaller blue regions have a thermal occupation of $\beta$ and $-\beta$, respectively, while the white regions separating them indicate maximally mixed states. Similar features (albeit at shifted temperatures) would be observed if the probe qubits were not on resonance.}
\end{figure}

A more comprehensive view of the stationary state of the system is obtained by plotting the mean spin value $\la \oJ_z \ra = (\chi-1)/2(\chi+1)$ as a function of the interaction time $\tau$ and of the ratio of coupling constants $g_y/g_x$ (Fig.~\ref{fig:XYint}). We have chosen some specific values for the other parameters, but we have checked that the qualitative behavior is stable over a wide range of values.

We see that for $g_y/g_x>0$ the steady state is very close to the Gibbs state for almost all values of $\tau$: this is because the RWA is usually valid and therefore the effective interaction is an exchange interaction. However, even for large values of $\tau$ we still find the reversed Gibbs state in narrow lines close to the condition $K=0$. This shows that a naive RWA stripping of the counter-rotating terms in $\oH_{\rm int}$ is not always valid even for long interaction times. Conversely, for $g_y/g_x<0$, the inverted Gibbs state is obtained in a wide range of parameters, especially for small $\tau$. The regions of thermalisation and anti-thermalisation are separated by a band in which $\la \oJ_z \ra\approx 0$ i.e.~$\chi\approx 1$. There, the steady state is maximally mixed: in thermalisation language, the channel acts as a bath of infinite temperature for the system.

\subsection{Product interaction ($g_y=g_z=0$)}\label{sec:product}

\begin{figure}
\centerline{\includegraphics[width=\columnwidth]{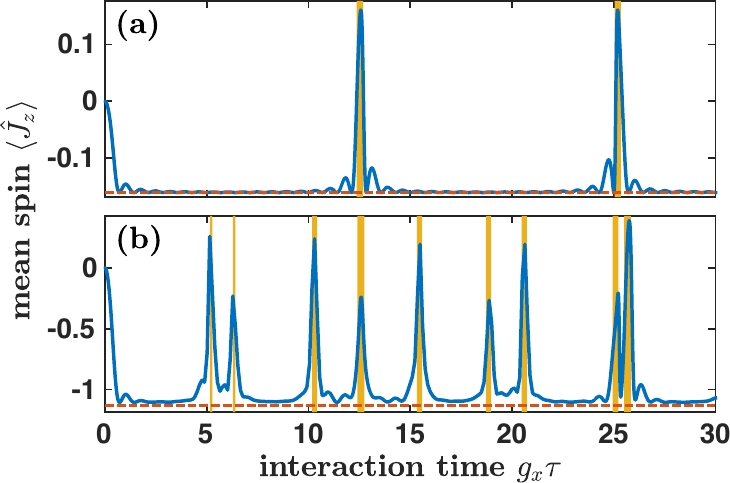}}
\caption{\label{fig:EJzXint} Steady state $\la \oJ_z \ra$ for (a) $J = 1/2$ and (b) $J = 2$ at different interaction times $\tau $ at a rate $\gamma =10^{-3}g_x $ through linear X-interaction with $\omega_p=4.4g_x,\,\,g_{y,z}=0$ using thermal resonant qubit probes $\omega_p = \omega_s$ at $k_BT = 1.5 \hbar\omega_p$. The dashed lines mark the thermal occupation at $\beta$ while the states are non-passive in the shaded regions.}
\end{figure}

In this subsection we study the instance $g_y=0$, i.e.~a product X-type interaction between system and probes, involving both resonant energy exchange and counter-rotating terms. The main features are already visible in the line $g_y/g_x=0$ of Fig.~\ref{fig:XYint}, which we plot on a linear scale as Fig.~\ref{fig:EJzXint}(a). But we extend our considerations to other spins than qubits [see Fig.~\ref{fig:EJzXint}(b)] and without imposing resonance.

In the short-time limit, both co- and counter-rotating coupling terms have equal contributions. The system equilibrates to a maximally mixed state with $\la \oJ_z\ra \to 0$. The effect of the system-probe interactions is then equivalent to that of a non-selective measurement process \cite{jacobs2006,wiseman2009} in the x-basis, which can be used as an entropy source for heat engines \cite{elouard2017,elouard2017maxwell,yi2017,elouard2018,buffoni2018}. We shall come back to this interpretation in Section \ref{sec:meas}. At longer interaction times, the steady state typically equilibrates close to a Gibbs state of inverse temperature $\beta_s = \beta \omega_p/\omega_s$, with occasional windows where the system could achieve non-passive steady states, i.e.~states with ergotropy \cite{allah2004work,goold2016review}. These windows are generally broader at strong couplings $g_x$ and more frequent for higher system spins.

\begin{figure}
\centerline{\includegraphics[width=\columnwidth]{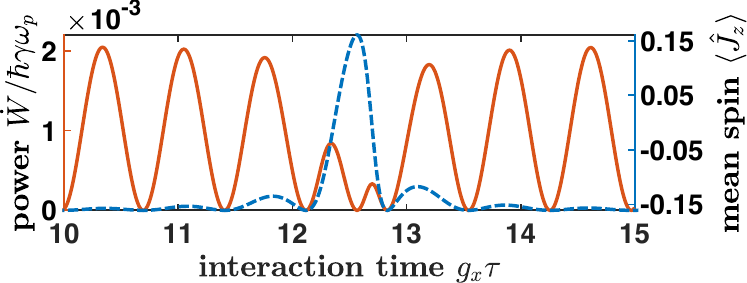}}
\caption{\label{fig:workXint} The dashed line is the mean spin $\la \oJ_z\ra$ from Fig.~\ref{fig:EJzXint}(a), zooming in a window near a population inversion point. The solid line is the corresponding steady-state work power \eqref{eq:workPower}, vanishing when the state is thermal as expected, but also when the state is anti-thermal ($g_x\tau\approx 12.5$).}
\end{figure}

If the system and the probes are qubits, the physics is rather simple and we have already sketched it above. When $\tau = 4\pi n/g_x$, the interaction is a partial swap between the two two-qubit states $|\frac{1}{2}, \frac{1}{2}\ra$ and $| -\frac{1}{2}, -\frac{1}{2}\ra $ modulo phases, which effectively describes a system interacting with an inverted thermal probe. At the more frequent values $\tau = 4\pi n/\sqrt{g_x^2 + 16\omega_s^2}$, the counter-rotating contributions cancel and the system equilibrates to a Gibbs state. Let us add here that the steady-state work power \eqref{eq:workPower} has an interesting behavior: besides vanishing at the thermal operation points as expected, it is suppressed in the vicinity of inversion points and is exactly zero at those points (Fig.~\ref{fig:workXint}). In other words, no work is needed to maintain the inverted thermal state, even though $\oU(\tau)$ is not an energy-preserving operation there. Of course, work must be spent to bring an initial state to the steady state: the repeated interaction process can be viewed as an ergotropy reservoir continuously `charging' the system qubit.

For higher system spins $J$ and still qubit probes, the interaction-induced energy level splitting leads to incommensurate frequencies; so one cannot find points in which the steady states are thermal or anti-thermal, even when $\la \oJ_z\ra$ is close to the value expected for such states [Fig.~\ref{fig:EJzXint}(b)]. The regions of population inversion now correspond to $\tau$-values describing a vanishing net resonant exchange  $|m, \frac{1}{2} \ra \leftrightarrow |m+1, -\frac{1}{2} \ra$ for $m<J$.

\subsection{Study of $g_y=\pm g_x$}

Moving away from the product interaction towards the top end ($g_x=g_y$) or the bottom end ($g_x=-g_y$) of Fig.~\ref{fig:XYint}, we arrive at the complementary limiting cases where the system always equilibrates either at the positive or negative probe temperature, regardless of the interaction time $\tau$. This follows from the formula \eqref{eq:qubitRatioGen} for the ratio of steady-state populations. 

For $g_x=g_y=g$ and $\omega_s=\omega_p$, the system-probe interaction describes a resonant exchange of excitation that preserves the total energy. In other words, the resonant probes realize a channel of thermal operations that effectively models spin thermalization, as often noticed and exploited \cite{scarani2002,landi2014,lorenzo2015,barra2015,hardal2015superradiant,lorenzo2015,manzano2017,Strasberg2017,Baumer2017,Barra2018,Hanson2018,grimmer2016,Grimmer2018,Grimmer2018b,deChiara2018,Pezzutto2018}. This holds true for arbitrary system spin numbers $J$, see App.~\ref{app:spinME}, as well as for harmonic oscillators. Should the probes be off-resonance, $\omega_p \neq \omega_s$, then the system will equilibrate to a Gibbs state of a different temperature from its environment, $\beta_s = \beta \omega_p/\omega_s$, as can be confirmed by noticing that $[(\omega_p/\omega_s)\oH_s + \oH_p,\oS]=0$. For qubits, $\oS$ reduces to a partial or full swap \cite{scarani2002,skrzypczyk2011,brask2015entangle}. 

Notice however that the master equation obtained from \eqref{eq:SmatrixME}, derived in App.~\ref{app:spinME}, is generally not equivalent to the standard Born-Markov spin thermalization model with Lindblad operators $\oJ_{\pm}$. Moreover, the process may be accompanied by additional dephasing when $g_z \neq 0$. In fact, the standard thermalisation master equation is retrieved only in the limiting scenario of short-time interactions ($g\tau \ll 1$) and for $\omega_s=\omega_p$ and $g_z=0$.

This has immediate implications when describing a system subjected to repeated exchange interactions in a thermal environment. Thermalization does \emph{not} emerge as a natural consequence even at prolonged interaction times with a broad spectrum of thermal probes in \eqref{eq:SmatrixMEfluct}, be it spins or harmonic oscillators --- contrary to the secular weak-coupling treatment with thermal oscillator baths. The reason lies in the inherent time dependence of the repeated interaction framework. Whenever the system and probes are off-resonance, the repeated switching of the interaction results in a non-vanishing work power \eqref{eq:workPower} that must come from external degrees of freedom. 

\begin{figure}
\centerline{\includegraphics[width=\columnwidth]{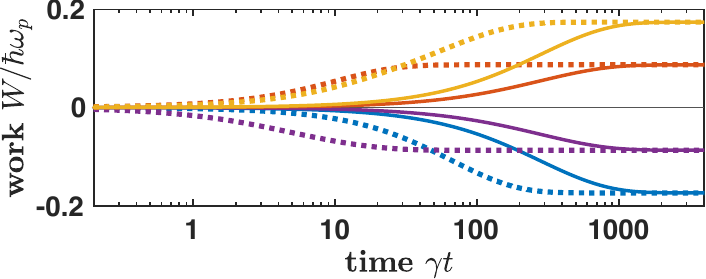}}
\caption{\label{fig:spinEquilWork} (color online) Accumulated work cost for a spin-$2$ system, initialised in its ground state, to reach steady state through dipole-dipole interaction $g_x=g_y=g_z=g$ with off-resonant qubit probes at temperature $k_BT = 1.5\hbar\omega_p$. We consider moderately strong dipole-dipole couplings with $g=0.05\omega_p$ at an average rate $\gamma = 10^{-3}\omega_p$. Solid and dotted lines correspond to short ($\omega_p \tau=2$) and long ($\omega_p \tau=200$) interaction times, respectively. The system is detuned from resonance with the probes according to $\omega_s/\omega_p = 0.8$ (blue, bottom), $0.9$ (purple), $1.1$ (red), and $1.2$ (yellow, top).}
\end{figure}

As an exemplary plot, we show in Fig.~\ref{fig:spinEquilWork} the cumulative work $W$ for dipole-dipole interactions for a system with $J=2$ as a function of time, for various system-probe detunings and interaction times. In all cases, the system eventually reaches the same steady state $\rho_{\infty} \sim \exp[-\beta \hbar \omega_p \oJ_z]$ and the work saturates, but the time to get there varies with $\tau$ and $g$, and the accumulated work cost depends on the detuning.

\subsection{A modified model: equilibration in a composite system}\label{sec:composite}

In the standard derivation of a thermalization master equation considering a composite system coupled weakly and simultaneously to a broad spectrum of environmental modes \cite{Breuer2002}, previous works have shown that the resulting master equation depends on the internal coupling and can describe either \emph{local} thermalization of the subsystem or thermalization to a \emph{global} Gibbs state \cite{rivas2010,levy2014,hofer2017,gonzalez2017,cresser2017coarse,stella2018fridge}.

In this subsection, we show that in the case of local repeated interactions with a composite system, we will only obtain a local master equation. 

Consider two linearly coupled spins,
\begin{eqnarray}
\oH_s = \hbar \omega_1 \oJ_z^{(1)} + \hbar \omega_2 \oJ_z^{(2)} + \hbar \sum_{k} G_k \oJ_k^{(1)}\,, \oJ_k^{(2)},\end{eqnarray} 
only one of which also interacts linearly with thermal probe spins $\oH_{\rm int} = \hbar \sum_{k} g_k \oJ_k^{(1)} \oj_k$. Contrary to the predictions of a secular master equation in the Born-Markov setting \cite{rivas2010,levy2014,hofer2017,gonzalez2017,cresser2017coarse}, this process will generally not bring the system to a global Gibbs state $\rho_{\infty}^{(12)} \propto \exp[-\beta \oH_s]$ regardless of the internal coupling strengths $G_k$ or the chosen probe frequency. A simple argument can be made based on the fact that $[\oH_0, \oH_{\rm int}] \neq 0$ as long as there is a non-vanishing internal coupling.
Hence, local repeated probe interactions cannot be made a thermal operation according to resource theory. 

Local equilibration, on the other hand, can be achieved with exchange interactions ($g_x=g_y$, $G_x=G_y$) as in the single-spin case \cite{barra2015,deChiara2018}. Specifically, the short-time scattering operator \eqref{eq:SmatrixTO} is precisely that of a single spin interacting with a probe as before, and so the corresponding dissipator in the master equation \eqref{eq:SmatrixME} yields a local Gibbs state 
in that limit. At finite $\tau$-values, indirect coupling contributions between $\ovJ^{(2)}$ and the probe start to appear in the scattering operator \eqref{eq:Smatrix}, implying that \eqref{eq:SmatrixME} does not resemble the local thermalization master equation. Nevertheless, the product Gibbs state rescaled to the probe occupation, $\rho_{\infty} \propto \exp[-\beta \hbar \omega_p (\oJ_z^{(1)} + \oJ_z^{(2)})]$, will remain a steady state of the repeated interaction process.

\section{Equilibration under indirect measurements}
\label{sec:meas}

This last section is motivated by the recent interest in the connection between thermodynamics and measurement \cite{elouard2017,elouard2017maxwell,yi2017,elouard2018,buffoni2018}. Hamiltonians of the type $\oH_{\rm{int}} = \hbar g\oA\otimes\oA_p$ can famously be read as describing the \textit{indirect measurement} of the observable $\oA$ of the system by changing the state of the probe (usually called \textit{pointer} in this context).

\begin{figure}
\centerline{\includegraphics[width=\columnwidth]{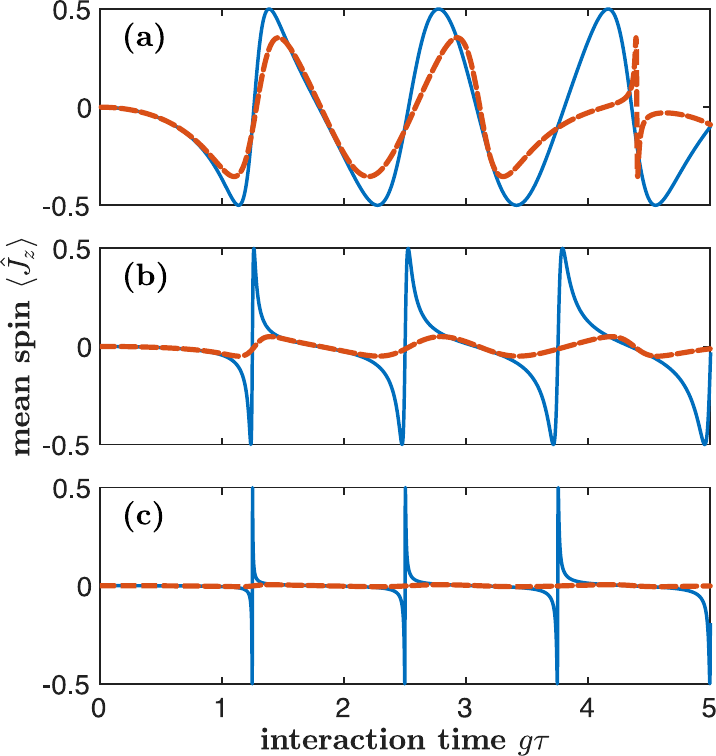}}
\caption{\label{fig:measurement} Steady state $\la \oJ_z \ra$ for a $J=1/2$ system of frequency $\omega_s=5g$ interacting with a probe qubit initialized at ground state with frequencies $\omega_p/\omega_s$ (a) $0.1$, (b) $0.01$ and (c) $0.001$ via a measurement interaction \eqref{eq:measurementH} where $\theta = \pi/100$ (dashed) and $\theta = \pi/2$ (solid) at a rate $\gamma =10^{-3}g $. }
\end{figure}

For definiteness, we consider two qubits with $\oH_s+\oH_p=\hbar \omega_s \oJ_z+\hbar \omega_p \oj_z$ as above, and with interaction Hamiltonian
\begin{equation}\label{eq:measurementH}
    \oH_{\rm{int}} = \hbar g (\cos\theta \oJ_z + \sin\theta\oJ_x)\otimes\oj_x .
\end{equation}
In each interaction, the pointer qubit (initialised in the ground state of $\oH_p$) is rotated around the $x$-axis clockwise or counterclockwise depending on the $\oA$-state of the system qubit. Whether each measurement is weak or strong is determined by the coupling strength $g$ and the interaction time $\tau$. In our repeated interaction framework, probing happens at random times, and between these times the system evolves under $\oH_s$. Thus, each pointer does not find the state left by the previous pointer, but a rotated version thereof, unless $[\oA,\oH_s]=0$. Under these conditions, we expect the repetition of ideal measurements to leave the system in the maximally mixed state (admittedly, such a repeated interaction channel is not a model of an informative measurement). This is indeed what we find for $\omega_p=0$, which is the ideal case for a pointer, insofar as its rotation and resetting do not require any investment of energy. 
Specifically, for $\omega_p=0$, the resulting master equation (App.~\ref{app:measurement}) contains two Lindblad operators of the form $\cos\alpha_k \oJ_z + \sin\alpha_k \oJ_x$, which generically align along different axes. Such dissipators are typically obtained in standard weak measurement master equations from expanding Gaussian POVMs along a measurement axis \cite{cresser2006,jacobs2006}. Since the Lindblad operators are Hermitian, the fixed point of the master equation is the maximally mixed state. Only when $[\oA,\oH_s]=0$, that is when $\oA=\oJ_z$, both dissipators become $\oJ_z$ and \emph{any} mixture of energy eigenstates is a steady state. In the case $\omega_p>0$, the steady state value of $\la \oJ_z \ra$ as a function of $\tau$ is plotted in Fig.~\ref{fig:measurement}. We see that the steady states remain close to the maximally mixed state ($\la \oJ_z \ra=0$) only for $\omega_p\ll\omega_s$.

\begin{figure}
\centerline{\includegraphics[width=\columnwidth]{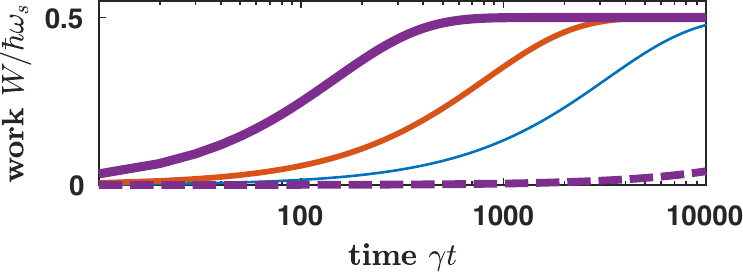}}
\caption{\label{fig:measureWork} (color online) Cumulative work with time for $\theta = \pi/2$ (solid lines) and $\theta = \pi/100$ (dashed line) considering different system-probe interaction times $g\tau=0.05$ (blue, thinnest), $0.1$ (red) and 1 (purple, thickest) at a rate $\gamma =10^{-3}g$ for an ideal probe with $\omega_p = 0$ interacting with a system initialized in the ground state.}
\end{figure}

Going back to $\omega_p=0$ and barring the case $[\oA,\oH_s]=0$, we have seen that the repeated interaction channel consisting of ideal measurements leads to unbounded increase of entropy. This is analogous to an infinite-temperature bath, and indeed the energy exchange by measurement is also sometimes termed `quantum heat' \cite{elouard2017}. However, similar to the observation made in  Ref.~\cite{strasberg2018}, here this `quantum heat' is actually work, originating in the switching on and off of $\oH_{\rm{int}}$. Indeed, referring to \eqref{eq:workPower}, a measurement of $\oA$ can increase the energy of the system when it does not commute with the free system Hamiltonian $\oH_0 = \oH_s$. In Fig.~\ref{fig:measureWork}, we show the cumulative work injected into a system initialized in the ground state. At any given time, the work is much lower when measurement is performed close to the z-axis (dashed) rather than along the x-axis (solid), in which case the energy of the system is significantly modified. Since $\omega_p=0$, the total work invested to get to the maximally mixed state should match the total change in energy of the system $W=\hbar\omega_s/2$. We see also that the cumulative work depends on the interaction time: while the total work invested to reach steady state should be the same, the rate at which work is injected is lower for weaker measurement.

\section{Conclusions}\label{sec:conclusion}

We have formulated a master equation to describe the average dynamics of a repeated interaction process between a system and a stream of probes. By adopting a scattering approach which coarse-grains over individual interactions and separates them from the free system evolution, our model gives accurate predictions across arbitrary interaction durations and strengths. 

As a testbed, we first considered finite-time repeated interactions with thermal probes. It turns out that such processes generally do not describe system thermalization to the environment temperature, except in tailored scenarios of resonant energy exchange between a single-gapped system and probes. Hence repeated interactions are incompatible with thermal operations once we consider off-resonant exchanges or local interactions within a composite system. Consequently, we typically attain out-of-equilibrium steady states in a thermal environment, including Gibbs-like states (albeit at ``wrong" temperatures), infinite-temperature and population-inverted states, depending on not just the form of interaction but also the interaction duration. 

We also modelled indirect measurements by pointer probes via repeated interactions and saw that an ideal measurement process by a gapless probe generally leads to infinite heating, apart from measurements that commute with the system Hamiltonian. This apparent ``heating'' is attributed to the work associated with turning on and off the interaction Hamiltonian.

Our formalism extends beyond the presented examples of thermalization and measurement and sets forth repeated interactions as a dynamical model that can be viewed as a tunable incoherent reservoir.

\section*{Acknowledgments}

We thank Daniel Grimmer, G\'eraldine Haack and Mart\'i Perarnau-Llobet for helpful discussions. SS thanks KITP for hospitality. This research is supported by the Singapore Ministry of Education through the Academic Research Fund Tier 3 (Grant No. MOE2012-T3-1-009); and by the same MoE and the National Research Foundation, Prime Minister's Office, Singapore, under the Research Centres of Excellence programme, as well as the National Science Foundation, USA (Grant No. NSF PHY-1748958).

%\clearpage 

\appendix 

\section{Stochastic simulations}\label{app:simulation}

To confirm the validity of our coarse-grained scattering master equations 
underlying the presented results, 
\eqref{eq:SmatrixME} and \eqref{eq:SmatrixMEfluct} in the main text, 
we performed Monte-Carlo simulations of the corresponding Poisson processes. 
Assuming that the system interacts with at most one probe at a time, the time evolution of a single trajectory is iterated as follows: We first draw a waiting time $\Delta t$ from the exponential distribution $p(\Delta t) = \gamma e^{-\gamma \Delta t}$ and evolve the system state $\rho_0 \to \rho_1$ freely for the time $\Delta t - \tau/2$. Now we switch on the interaction and attach the thermal probe state $\eta$ to the system. If an ensemble of probes (e.g.~of different energies) is considered, in \eqref{eq:SmatrixMEfluct}, 
we could draw the probe parameters from the given distribution $p(\xi)$. The combined density matrix $\rho_1 \otimes \eta$ is then evolved for the time $\tau$ according to the unitary $\oU (\tau)$ given in the main text. After tracing out the probe, we proceed to the next iteration with the reduced system state, drawing a new waiting time, etc. The state is stored at fixed times and averaged over $N \gg 1$ trajectories. 

\begin{figure*}[h]
\centerline{\includegraphics[width=0.9\textwidth]{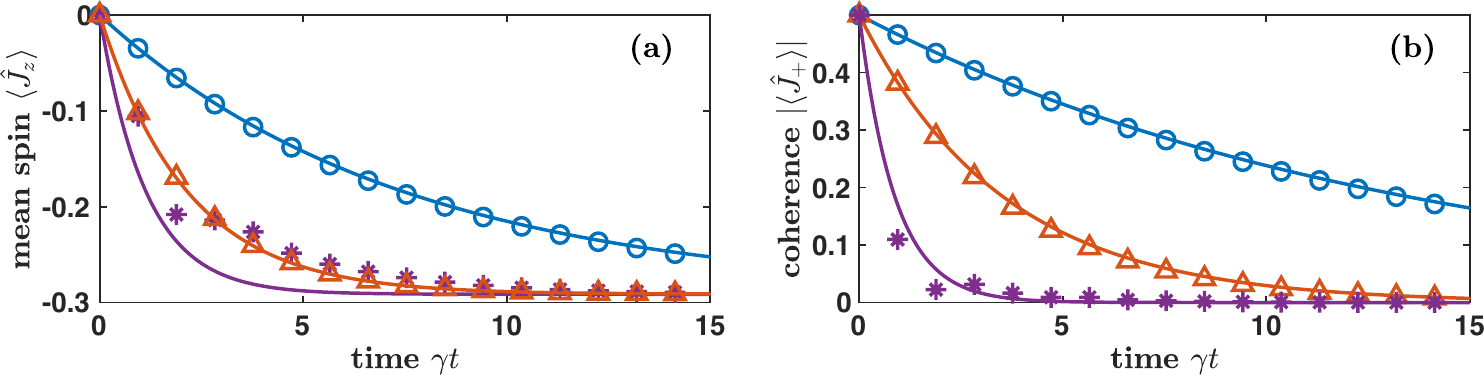}}
\caption{\label{fig:compSimME} (color online) Time evolution of (a) the mean spin $\la \oJ_z \ra$ and (b) the coherence $| \la \oJ_+ \ra |$.
We consider a qubit system initialized at $\ket{+}$ interacting with resonant thermal qubit probes at $k_BT = 0.75 \hbar\omega_p$, $(g_x,g_y,g_z) = (5,2.5,1) \times 10^{-2}\omega_p$, and $\gamma = 2.5 \times 10^{-3}\omega_p$.
The stochastic simulation results averaged over $10^6$ trials (markers) are compared to the predictions of the ensemble-averaged master equation (lines) for $\gamma \tau= 0.05$ (blue, circles), $0.1$ (red, triangles) and $1$ (purple, stars).} 
\end{figure*}

\begin{figure*}[h]
\centerline{\includegraphics[width=0.9\textwidth]{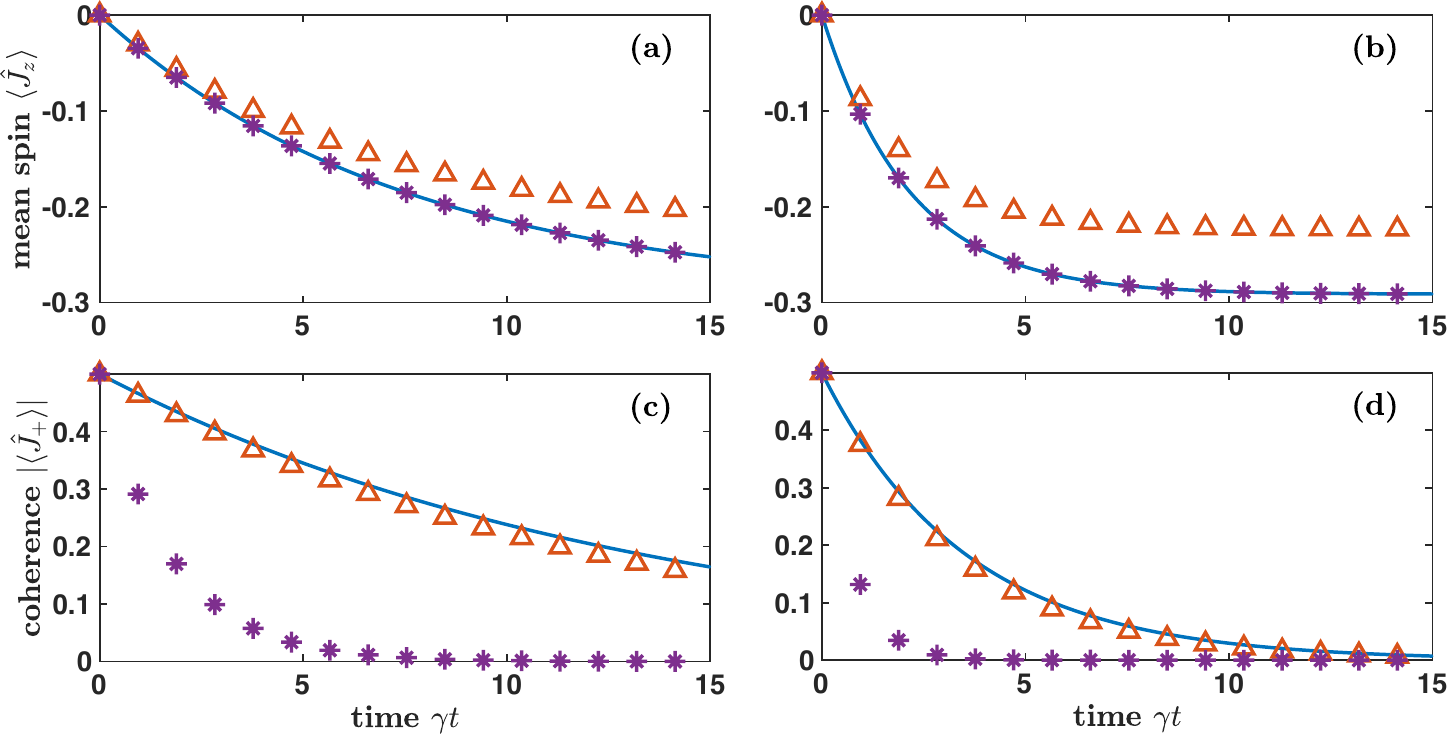}}
\caption{\label{fig:compSUSeik} We compare the master equations derived using $\oS$ (solid lines) with $\oU (\tau)$ (stars) and $\oS_{\rm eik}$ (triangles) using the same parameter settings as Fig.~\ref{fig:compSimME}. The left panels (a) and (c) correspond to $\gamma \tau= 0.05$, (b) and (d) to $0.1$. } 
\end{figure*}

Generally, we found excellent agreement between the averaged numerical results and the predictions of the master equation for $\gamma \tau < 0.1$. At larger $\gamma \tau$-values, the coarse-grained Poisson model of scattering events ceases to be valid.
This is illustrated in Fig.~\ref{fig:compSimME}, where we compare the results for a qubit interacting with resonant qubit probes in terms of (a) its mean spin and (b) magnitude of coherence. 

To further illustrate the validity and significance of our scattering-based method for finite-time interactions, suppose we replace $\oS$ in our master equation \eqref{eq:SmatrixME} 
by the short-time operator $\oS_{\rm eik}$ (which depends only on the system-probe interaction Hamiltonian) or the unitary map $\oU(\tau)$ (which represents the combined system-probe time evolution). The latter could arise if a `reset' approach as in \cite{lidar2001} would be employed to model the present finite-time repeated interaction process, for instance. 
Figure \ref{fig:compSUSeik} shows that these na\"{i}ve implementations of the time evolution predict results that do not match our master equation model (solid lines) and the ensemble-averaged random process. In particular, while $\oU(\tau)$ (stars) reproduces the evolution of the spin population in (a) and (b), it overestimates the decoherence effect in (c) and (d). The reason lies in an additional dephasing caused by the contribution of the free Hamiltonian to the time evolution described by $\oU(\tau)$; this unphysical dephasing would be present even for $g_{x,y,z} \to 0$. 

\begin{widetext}

\section{Master equation for linear interactions between qubit system and probes}\label{app:XXME}

We derive the master equation for the linear interaction $\oH_{\rm int} = \hbar \sum_k g_k \oJ_k \otimes \oj_k$ 
for $J = j = 1/2$ assuming that the probes are prepared in the Gibbs state $\eta$ at inverse temperature $\beta$. The unitary that corresponds to the free system-probe evolution is given by
\begin{equation}\label{eq:XXfreeU}
\oU_0(\tau) = e^{i\Omega \tau/2} \ketbra{-\half,-\half}{-\half,-\half}  + e^{-i\Omega \tau/2} \ketbra{\half,\half}{\half,\half} + e^{i\Delta \tau/2} \ketbra{-\half,\half}{-\half,\half}  + e^{-i\Delta \tau/2} \ketbra{\half,-\half}{\half,-\half},
\end{equation}
where $\Omega = \omega_s + \omega_p$ and $\Delta = \omega_s - \omega_p$. 
In this case, $\oH = \oH_0 + \oH_{\rm int}$ is block-diagonal in subspaces $\left\{ \ket{-\half,-\half}, \ket{\half,\half} \right\}$ and $\left\{\ket{-\half,\half},\ket{\half,-\half}\right\}$,
\begin{eqnarray}
\oH = \frac{\hbar}{4}
 \begin{bmatrix}
  -2\Omega + g_z & 0 & 0 & g_x-g_y \\
  0 & -2\Delta - g_z  & g_x+g_y & 0\\
  0   & g_x+g_y & 2\Delta - g_z & 0\\
  g_x-g_y & 0 & 0 & 2\Omega + g_z
 \end{bmatrix}\nonumber\\.
\end{eqnarray}
Introducing $G_\Delta = (g_x + g_y)/2$ and $G_\Omega = (g_x - g_y)/2$, the eigenvalues and eigenvectors are
\begin{eqnarray}\label{eq:XXeig}
\lambda_\Delta^\pm &=& \frac{\hbar}{4}\left(-g_z\pm 2 \sqrt{G_\Delta^2 + \Delta^2}\right), \quad \lambda_\Omega^\pm = \frac{\hbar}{4}\left(g_z\pm 2 \sqrt{G_\Omega^2 + \Omega^2}\right), \nonumber\\
\ket{\Delta^+} &\propto& -G_\Delta \ket{\half,-\half} + \left(\Delta + \sqrt{G_\Delta^2 + \Delta^2}\right) \ket{-\half,\half}, \nonumber\\
\ket{\Delta^-} &\propto & \left(\Delta + \sqrt{G_\Delta^2 + \Delta^2}\right)  \ket{\half,-\half} + G_\Delta\ket{-\half,\half}, \nonumber\\
\ket{\Omega^+} &\propto& \left(\Omega + \sqrt{G_\Omega^2 + \Omega^2}\right)  \ket{\half,\half} + G_\Omega\ket{-\half,-\half}, \nonumber\\
\ket{\Omega^-} &\propto& -G_\Omega \ket{\half,\half} + \left(\Omega + \sqrt{G_\Omega^2 + \Omega^2}\right) \ket{-\half,-\half}.
\end{eqnarray}
Then the unitaries $\oU(\tau)$ and $\oS$ that describe the scattering event are given by
\begin{eqnarray}\label{eq:XXUfull}
\oU(\tau) &=& \sum_{\omega=\Omega,\Delta} e^{-i\lambda_\omega^+\tau/\hbar} \ketbra{\omega^+}{\omega^+} +  e^{-i\lambda_\omega^-\tau/\hbar} \ketbra{\omega^-}{\omega^-}. \\
\oS &=& e^{-ig_z\tau/4}\left\{C_\Omega \ketbra{-\half,-\half}{-\half,-\half} + C_\Omega^* \ketbra{\half,\half}{\half,\half}  - iK_\Omega \left(\ketbra{-\half,-\half}{\half,\half} + h.c.\right) \right\}\nonumber\\
&&+ e^{ig_z\tau/4}\left\{[C_\Delta \ketbra{-\half,\half}{-\half,\half} + C_\Delta^* \ketbra{\half,-\half}{\half,-\half} - iK_\Delta \left(\ketbra{-\half,\half}{\half,-\half} + h.c. \right)\right\}.
\end{eqnarray}
Here we introduce
\begin{eqnarray}
C_\omega &=& e^{-i\omega\tau/2}\, \frac{ e^{i\sqrt{{G_\omega^2 + \omega^2}}\tau/2} \left(\omega+\sqrt{G_\omega^2 + \omega^2}\right)^2 + e^{-i\sqrt{{G_\omega^2 + \omega^2}}\tau/2} G_\omega^2 }{G_\omega^2 + \left(\omega+\sqrt{G_\omega^2 + \omega^2}\right)^2} , \nonumber \\
K_\omega &=& \frac{2G_\omega\left(\omega+\sqrt{G_\omega^2 + \omega^2}\right) }{G_\omega^2 + \left(\omega+\sqrt{G_\omega^2 + \omega^2}\right)^2}\sin\frac{\sqrt{G_\omega^2 + \omega^2}}{2}\tau.
\end{eqnarray}
The master equation is then given by
\begin{eqnarray}
\dot{\rho} &=& -i\left( \omega_s+ \gamma \frac{e^{\beta\omega_p/2}\Im \{e^{-ig_z\tau/2 }C_\Omega C_\Delta\}+ e^{-\beta\omega_p/2}\Im \{e^{ig_z\tau/2 }C_\Omega C_\Delta\} }{Z_p} \right) [\oJ_z,\rho]\nonumber\\ && + \frac{\gamma}{Z_p}\left(e^{\beta\omega_p/2} \left|C_\Omega e^{-ig_z\tau/4}-C_\Delta^* e^{ig_z\tau/4}\right|^2+e^{-\beta\omega_p/2} \left|C_\Omega e^{ig_z\tau/4}-C_\Delta^* e^{-ig_z\tau/4}\right|^2\right)\cD[\oJ_z]\rho\nonumber\\ &&+ \frac{\gamma}{Z_p}\left(e^{\beta\omega_p/2} K_\Omega^2 + e^{-\beta\omega_p/2} K_\Delta^2 - Z_p K_\Omega K_\Delta\right)\cD [\oJ_+]\rho + \frac{\gamma}{Z_p}\left(e^{\beta\omega_p/2} K_\Delta^2 + e^{-\beta\omega_p/2} K_\Omega^2 - Z_p K_\Omega K_\Delta\right)\cD [\oJ_-]\rho\nonumber\\ &&+ \frac{4\gamma K_\Omega K_\Delta}{Z_p}\left(e^{\beta\omega_p/2} \cD \left[\cos{\frac{g_z\tau}{4}}\oJ_x + \sin\frac{g_z\tau}{4}\oJ_y\right]\rho + e^{-\beta\omega_p/2} \cD \left[\cos{\frac{g_z\tau}{4}}\oJ_x - \sin\frac{g_z\tau}{4}\oJ_y\right]\rho\right).
\end{eqnarray}
At steady state, the ratio of the probability of excited and ground states is
\begin{equation}\label{eq:popratio}
\chi = \frac{\bra{\half}\rho\ket{\half}}{\bra{-\half}\rho\ket{-\half}} = \frac{e^{\beta\omega_p/2} K_\Omega^2 + e^{-\beta\omega_p/2} K_\Delta^2}{e^{\beta\omega_p/2} K_\Delta^2 + e^{-\beta\omega_p/2} K_\Omega^2}. 
\end{equation}
In particular, if one either sets $g_x=-g_y$ or chooses $\tau$ such that $K_\Delta = 0$, i.e. $\sqrt{G_\Delta^2 + \Delta^2} \tau = 2n\pi$, then \eqref{eq:popratio} reduces to 
$\chi = e^{\beta\omega_p}$, i.e.~ the steady state is a negative-temperature state with $\beta_s = -\beta\omega_p/\omega_s$. 
For exchange interactions ($g_x=g_y$), the steady state will be a Gibbs state at $\beta_s = \beta\omega_p/\omega_s$. 
In general, \eqref{eq:popratio} reduces to 
\begin{equation}\label{eq:popratioShortTime}
\chi \approx 1 - \frac{4g_xg_y\sinh{\left(\beta\omega_p/2\right)}}{(g_x^2 + g_y^2)\cosh{\left(\beta\omega_p/2\right)} + 2g_xg_y \sinh{\left(\beta\omega_p/2\right)}} 
\end{equation}
in the short-time limit where $\tau\ll \omega_{s,p}^{-1},\, g_{x,y}^{-1} $. When $g_x$ or $g_y = 0$, 
the system would be maximally mixed, i.e.~ \eqref{eq:popratioShortTime} reduces to $\chi \approx 1$. This agrees with the short-time behavior of product interactions discussed in the main text.

\section{Master equation for exchange interaction between a spin-J system and spin-1/2 probes}\label{app:spinME}

Here we give an explicit form of the repeated interaction master equation \eqref{eq:SmatrixME} 
for the case of exchange spin-qubit interactions ($g_x = g_y= g$). For the system spin, we use dimensionless spin operators and define the usual $(2J+1)$-dimensional algebra via
\begin{equation}
[\oJ_k,\oJ_\ell] = i\varepsilon_{k\ell n} \oJ_n, \quad \oJ_{\pm} = \oJ_x \pm i\oJ_y, \quad 
\oJ_z |m\ra = m|m\ra, \quad \ovJ^2 = J(J+1) \id .
\end{equation}
The unitary that corresponds to the free system-probe evolution is given by
\begin{equation}\label{eq:spinJfreeU}
\oU_0(\tau) = \sum_{m=-J}^{J} \left[ e^{-i(\omega_s m+\omega_p/2) \tau} \ketbra{m,\half}{m,\half} + e^{-i(\omega_s m-\omega_p/2) \tau} \ketbra{m,-\half}{m,-\half} \right].
\end{equation}
To obtain $\oU(\tau)$, we observe that $\oH = \oH_0 + \oH_{\rm int}$ is block-diagonal in the subspaces $\left\{\ket{m,-\half},\ket{m-1,\half}\right\}$ for $-J<m\leq J$, with eigenvalues and eigenvectors
\begin{eqnarray}\label{eq:spinJeig}
\lambda_m^\pm &=& \frac{\hbar}{4}\left[\left(m-\frac{1}{2}\right)\omega_s -g_z\pm 2 \sqrt{G_m^2 + \Delta_m^2}\right], \nonumber \\
\ket{\psi_m^+} &\propto& -G_m \ket{m,-\half} + \left(\Delta_m + \sqrt{G_m^2 + \Delta_m^2}\right) \ket{m-1,\half}, \nonumber\\
\ket{\psi_m^-} &\propto& \left(\Delta_m + \sqrt{G_m^2 + \Delta_m^2}\right)  \ket{m,-\half} + G_m\ket{m-1,\half}, 
\end{eqnarray}
where we define $\Delta_m = \Delta-\left(m-\half\right)g_z$ and $G_m = g\sqrt{J(J+1)-m(m-1)}/2$.
Then the unitary reads as
\begin{eqnarray}\label{eq:spinJUtotal}
\oU(\tau) &=& \!\! \sum_{m=-J+1}^{J} \left( e^{-i\lambda_m^+ \tau/\hbar} \ketbra{\psi_m^+}{\psi_m^+} + e^{-i\lambda_m^- \tau/\hbar} \ketbra{\psi_m^-}{\psi_m^-} \right) \nonumber \\
&& + e^{-i(2\omega_s J+\omega_p+g_zJ) \tau/2}\ketbra{J,\half}{J,\half} + e^{i(2\omega_s J+\omega_p-g_zJ) \tau/2}\ketbra{-J,-\half}{-J,-\half}.
\end{eqnarray}
The unitary scattering operator $\oS$ is
\begin{eqnarray}\label{eq:spinJS}
\oS &=& e^{ig_z\tau/4} \sum_{m=-J+1}^{J} \left[ C_m\ketbra{m,-\half}{m,-\half} + C_m^*\ketbra{m-1,\half}{m-1,\half} - i K_m \left(\ketbra{m,-\half}{m-1,\half} + h.c.\right) \right] \nonumber \\
&& + \, e^{ig_zJ\tau/2} \left( \ketbra{-J,-\half}{-J,-\half}  + \ketbra{J,\half}{J,\half} \right), 
\end{eqnarray}
where 
\begin{eqnarray}\label{eq:spinJCK}
C_m &=& e^{-i\Delta \tau/2} \, \frac{ e^{i\sqrt{{G_m^2 + \Delta_m^2}}\tau/2}\left(\Delta_m+\sqrt{G_m^2 + \Delta_m^2}\right)^2 + e^{-i\sqrt{{G_m^2 + \Delta_m^2}}\tau/2}G_m^2}{G_m^2 + \left(\Delta_m+\sqrt{G_m^2 + \Delta_m^2}\right)^2},\nonumber\\
K_m &=& \frac{2G_m\left(\Delta_m+\sqrt{G_m^2 + \Delta_m^2}\right) }{G_m^2 + \left(\Delta_m+\sqrt{G_m^2 + \Delta_m^2}\right)^2}\sin\frac{\sqrt{G_m^2 + \Delta_m^2}}{2}\tau.
\end{eqnarray}
Again, if the probes are prepared in the Gibbs state $\eta$ at inverse temperature $\beta$, an explicit master equation can be derived from \eqref{eq:SmatrixME}, 
\begin{equation}\label{eq:spinJME}
\dot{\rho} = -i\omega_s [\oJ_z,\rho] + \frac{\gamma e^{\beta\omega_p/2}}{Z_p} \cD [\oL_1]\rho + \frac{\gamma e^{-\beta\omega_p/2}}{Z_p} \cD [\oL_2]\rho +  \frac{\gamma e^{\beta\omega_p/2}}{Z_p} \cD [\oL_{-}]\rho + \frac{\gamma e^{-\beta\omega_p/2}}{Z_p} \cD [\oL_+]\rho, 
\end{equation}
with the Lindblad operators
\begin{eqnarray}
\oL_1 &=& e^{i g_z J\tau/2} \ketbra{-J}{-J} + e^{i g_z \tau/4}\sum_{m=-J+1}^{J} C_m^*\ketbra{m}{m},\nonumber\\
\oL_2 &=& e^{i g_z \tau/4}\sum_{m=-J}^{J-1} C_m\ketbra{m}{m} + e^{i g_z J\tau/2} \ketbra{J}{J},\nonumber\\
\oL_{+} & =& e^{i g_z \tau/4} \! \! \sum_{m=-J+1}^{J} K_m\ketbra{m}{m-1}, \quad \oL_{-} = \oL_{+}^\dg .
\end{eqnarray}
\eqref{eq:spinJME} predicts the decay of all coherences, as well as a Gibbs thermal state at temperature $\beta_s = \beta\omega_p/\omega_s$ since $\left\langle m-1|\rho_\infty | m-1 \right\rangle/\left\langle m|\rho_\infty | m \right\rangle = e^{\beta \omega_p}$ for $-J<m\leq J$. In addition, we see that for $J\neq 1/2$, the dissipators do \emph{not} reduce to $\oJ_\pm$, which implies that the thermalization rates would defer from the standard Born-Markov spin thermalization master equation
\begin{equation}\label{eq:standardME}
\dot{\rho}  = -i \omega_s\left[ \oJ_z,\rho\right] + \Gamma e^{\beta\omega_s/2} \cD [\oJ_-]\rho +\Gamma e^{-\beta\omega_s/2} \cD [\oJ_+]\rho,
\end{equation}
where $\Gamma$ is an arbitrary thermalization rate derived under this framework.

To illustrate these differences more explicitly, we consider the master equation of a spin-$1/2$ system
\begin{equation}\label{eq:spinhalfME}
\dot{\rho}  = -i\left(\omega_s + \gamma \Im \{C_{1/2} \} \right)\left[\oJ_z,\rho\right] + \gamma\left|1-C_{1/2}\right|^2 \cD [\oJ_z]\rho + \frac{\gamma |K_{1/2}|^2}{Z_p} \left( e^{\beta\omega_p/2} \cD [\oJ_-]\rho +e^{-\beta\omega_p/2} \cD [\oJ_+]\rho \right). 
\end{equation}
From \eqref{eq:spinhalfME}, we see that the repeated exchange interaction between the probes and the system not only leads to dissipation terms, but also to dephasing and a shift in energy. However, these latter terms are of higher order in $g$ and $\tau$ and therefore negligible in the short-time limit.

\section{Master equation for measurement interaction with ideal probe}\label{app:measurement}

We consider the repeated measurement master equation with interaction Hamiltonian \eqref{eq:measurementH} for an ideal measurement probe, i.e.~$\omega_p=0$.
The Hamiltonians $\oH_0 = \hbar\omega_s\oJ_z\otimes\mathbbm{1}$ and $\oH = \oH_0 + \hbar g (\cos\theta \oJ_z + \sin\theta\oJ_x)\otimes\oj_x$ are block-diagonal in the eigenbasis representation $\ket{\pm}$ of $\oj_x$,
\begin{equation}
\oH = \hat{M}_+ \otimes \ketbra{+}{+} + \hat{M}_- \otimes \ketbra{-}{-}, \qquad \hat{M}_{\pm }  = (\hbar\omega_s \pm \hbar g \cos\theta)\oJ_z \pm \hbar g \sin\theta\oJ_x .
\end{equation}
The scattering matrix $\oS$ can be written as $\oS = \sum_{j=\pm } \hat{K}_j \otimes \ketbra{j}{j}$ where 
\begin{eqnarray}
\hat{K}_{\pm }  &=& \ e^{i  \hbar\omega_s\oJ_z\tau/2} e^{-i  \hat{M}_{\pm }\tau} e^{i  \hbar\omega_s\oJ_z\tau/2} = A_\pm \mathbbm{1}+ 2iB_\pm \oJ_z \mp 2i C_\pm \oJ_x, \nonumber\\
A_\pm &=& \left[\cos\left(\frac{\omega_s \tau}{2}\right)\cos (R_\pm\tau) + \sin\left(\frac{\omega_s \tau}{2}\right) \frac{ \omega_s \pm 2g\cos\theta\sin(R_\pm\tau)}{2R_\pm}\right], \nonumber\\
B_\pm &=& \left[\sin\left(\frac{\omega_s \tau}{2}\right)\cos (R_\pm\tau) - \cos\left(\frac{\omega_s \tau}{2}\right) \frac{ \omega_s \pm 2g\cos\theta\sin(R_\pm\tau)}{2R_\pm}\right], \nonumber\\
C_\pm &=& \frac{ g\sin\theta\sin(R_\pm\tau)}{R_\pm}, \qquad 
R_{\pm} = \frac{\sqrt{4g^2+\omega_s^2 \pm 4g\omega_s\cos\theta}}{2}.
\end{eqnarray}
Assuming that the probe is initialized in the ground state, the master equation is given by
\begin{eqnarray}
        \dot{\rho} &=& -i\omega_s [\oJ_z,\rho] + \frac{\gamma}{2} \left(\hat{K}_{+}\rho\hat{K}^\dg_{+} + \hat{K}_{-}\rho\hat{K}^\dg _{-} -2\rho \right) \nonumber\\
    &=& -i\left[\omega_s \oJ_z - \gamma \left(A_+ B_+ + A_- B_- \right) \oJ_z + \gamma \left(A_- B_- - A_+ B_+ \right)\oJ_x,\rho\right] \nonumber\\
    &+& \gamma\sqrt{B_+^2 + C_+^2}\cD [B_+ \oJ_z + C_+ \oJ_x]\rho +\gamma\sqrt{B_-^2 + C_-^2} \cD [B_- \oJ_z - C_- \oJ_x]\rho,
\end{eqnarray}
which effectively describes weak measurements in the basis of $B_+\oJ_z + C_+ \oJ_x$ and $B_- \oJ_z - C_- \oJ_x$.
\end{widetext}

%\bibliography{addLit}

\begin{thebibliography}{50}%
\makeatletter
\providecommand \@ifxundefined [1]{%
 \@ifx{#1\undefined}
}%
\providecommand \@ifnum [1]{%
 \ifnum #1\expandafter \@firstoftwo
 \else \expandafter \@secondoftwo
 \fi
}%
\providecommand \@ifx [1]{%
 \ifx #1\expandafter \@firstoftwo
 \else \expandafter \@secondoftwo
 \fi
}%
\providecommand \natexlab [1]{#1}%
\providecommand \enquote  [1]{``#1''}%
\providecommand \bibnamefont  [1]{#1}%
\providecommand \bibfnamefont [1]{#1}%
\providecommand \citenamefont [1]{#1}%
\providecommand \href@noop [0]{\@secondoftwo}%
\providecommand \href [0]{\begingroup \@sanitize@url \@href}%
\providecommand \@href[1]{\@@startlink{#1}\@@href}%
\providecommand \@@href[1]{\endgroup#1\@@endlink}%
\providecommand \@sanitize@url [0]{\catcode `\\12\catcode `\$12\catcode
  `\&12\catcode `\#12\catcode `\^12\catcode `\_12\catcode `\%12\relax}%
\providecommand \@@startlink[1]{}%
\providecommand \@@endlink[0]{}%
\providecommand \url  [0]{\begingroup\@sanitize@url \@url }%
\providecommand \@url [1]{\endgroup\@href {#1}{\urlprefix }}%
\providecommand \urlprefix  [0]{URL }%
\providecommand \Eprint [0]{\href }%
\providecommand \doibase [0]{http://dx.doi.org/}%
\providecommand \selectlanguage [0]{\@gobble}%
\providecommand \bibinfo  [0]{\@secondoftwo}%
\providecommand \bibfield  [0]{\@secondoftwo}%
\providecommand \translation [1]{[#1]}%
\providecommand \BibitemOpen [0]{}%
\providecommand \bibitemStop [0]{}%
\providecommand \bibitemNoStop [0]{.\EOS\space}%
\providecommand \EOS [0]{\spacefactor3000\relax}%
\providecommand \BibitemShut  [1]{\csname bibitem#1\endcsname}%
\let\auto@bib@innerbib\@empty
%</preamble>
\bibitem [{\citenamefont {Scarani}\ \emph {et~al.}(2002)\citenamefont
  {Scarani}, \citenamefont {Ziman}, \citenamefont {\ifmmode \check{S}\else
  \v{S}\fi{}telmachovi\ifmmode~\check{c}\else \v{c}\fi{}}, \citenamefont
  {Gisin},\ and\ \citenamefont {Bu\ifmmode~\check{z}\else
  \v{z}\fi{}ek}}]{scarani2002}%
  \BibitemOpen
  \bibfield  {author} {\bibinfo {author} {\bibfnamefont {V.}~\bibnamefont
  {Scarani}}, \bibinfo {author} {\bibfnamefont {M.}~\bibnamefont {Ziman}},
  \bibinfo {author} {\bibfnamefont {P.}~\bibnamefont {\ifmmode \check{S}\else
  \v{S}\fi{}telmachovi\ifmmode~\check{c}\else \v{c}\fi{}}}, \bibinfo {author}
  {\bibfnamefont {N.}~\bibnamefont {Gisin}}, \ and\ \bibinfo {author}
  {\bibfnamefont {V.}~\bibnamefont {Bu\ifmmode~\check{z}\else \v{z}\fi{}ek}},\
  }\href {\doibase 10.1103/PhysRevLett.88.097905} {\bibfield  {journal}
  {\bibinfo  {journal} {Phys. Rev. Lett.}\ }\textbf {\bibinfo {volume} {88}},\
  \bibinfo {pages} {097905} (\bibinfo {year} {2002})}\BibitemShut {NoStop}%
\bibitem [{\citenamefont {Landi}\ \emph {et~al.}(2014)\citenamefont {Landi},
  \citenamefont {Novais}, \citenamefont {de~Oliveira},\ and\ \citenamefont
  {Karevski}}]{landi2014}%
  \BibitemOpen
  \bibfield  {author} {\bibinfo {author} {\bibfnamefont {G.~T.}\ \bibnamefont
  {Landi}}, \bibinfo {author} {\bibfnamefont {E.}~\bibnamefont {Novais}},
  \bibinfo {author} {\bibfnamefont {M.~J.}\ \bibnamefont {de~Oliveira}}, \ and\
  \bibinfo {author} {\bibfnamefont {D.}~\bibnamefont {Karevski}},\ }\href
  {\doibase 10.1103/PhysRevE.90.042142} {\bibfield  {journal} {\bibinfo
  {journal} {Phys. Rev. E}\ }\textbf {\bibinfo {volume} {90}},\ \bibinfo
  {pages} {042142} (\bibinfo {year} {2014})}\BibitemShut {NoStop}%
\bibitem [{\citenamefont {Lorenzo}\ \emph {et~al.}(2015)\citenamefont
  {Lorenzo}, \citenamefont {McCloskey}, \citenamefont {Ciccarello},
  \citenamefont {Paternostro},\ and\ \citenamefont {Palma}}]{lorenzo2015}%
  \BibitemOpen
  \bibfield  {author} {\bibinfo {author} {\bibfnamefont {S.}~\bibnamefont
  {Lorenzo}}, \bibinfo {author} {\bibfnamefont {R.}~\bibnamefont {McCloskey}},
  \bibinfo {author} {\bibfnamefont {F.}~\bibnamefont {Ciccarello}}, \bibinfo
  {author} {\bibfnamefont {M.}~\bibnamefont {Paternostro}}, \ and\ \bibinfo
  {author} {\bibfnamefont {G.~M.}\ \bibnamefont {Palma}},\ }\href {\doibase
  10.1103/PhysRevLett.115.120403} {\bibfield  {journal} {\bibinfo  {journal}
  {Phys. Rev. Lett.}\ }\textbf {\bibinfo {volume} {115}},\ \bibinfo {pages}
  {120403} (\bibinfo {year} {2015})}\BibitemShut {NoStop}%
\bibitem [{\citenamefont {Grimmer}\ \emph {et~al.}(2016)\citenamefont
  {Grimmer}, \citenamefont {Layden}, \citenamefont {Mann},\ and\ \citenamefont
  {Mart{\'{i}}n-Mart{\'{i}}nez}}]{grimmer2016}%
  \BibitemOpen
  \bibfield  {author} {\bibinfo {author} {\bibfnamefont {D.}~\bibnamefont
  {Grimmer}}, \bibinfo {author} {\bibfnamefont {D.}~\bibnamefont {Layden}},
  \bibinfo {author} {\bibfnamefont {R.~B.}\ \bibnamefont {Mann}}, \ and\
  \bibinfo {author} {\bibfnamefont {E.}~\bibnamefont
  {Mart{\'{i}}n-Mart{\'{i}}nez}},\ }\href {\doibase 10.1103/PhysRevA.94.032126}
  {\bibfield  {journal} {\bibinfo  {journal} {Phys. Rev. A}\ }\textbf {\bibinfo
  {volume} {94}},\ \bibinfo {pages} {032126} (\bibinfo {year}
  {2016})}\BibitemShut {NoStop}%
\bibitem [{\citenamefont {B{\"a}umer}\ \emph {et~al.}(2017)\citenamefont
  {B{\"a}umer}, \citenamefont {Perarnau-Llobet}, \citenamefont {Kammerlander},\
  and\ \citenamefont {Renner}}]{Baumer2017}%
  \BibitemOpen
  \bibfield  {author} {\bibinfo {author} {\bibfnamefont {E.}~\bibnamefont
  {B{\"a}umer}}, \bibinfo {author} {\bibfnamefont {M.}~\bibnamefont
  {Perarnau-Llobet}}, \bibinfo {author} {\bibfnamefont {P.}~\bibnamefont
  {Kammerlander}}, \ and\ \bibinfo {author} {\bibfnamefont {R.}~\bibnamefont
  {Renner}},\ }\href {https://arxiv.org/abs/1712.07128} {\bibfield  {journal}
  {\bibinfo  {journal} {arXiv:1712.07128}\ } (\bibinfo {year}
  {2017})}\BibitemShut {NoStop}%
\bibitem [{\citenamefont {Strasberg}\ \emph {et~al.}(2017)\citenamefont
  {Strasberg}, \citenamefont {Schaller}, \citenamefont {Brandes},\ and\
  \citenamefont {Esposito}}]{Strasberg2017}%
  \BibitemOpen
  \bibfield  {author} {\bibinfo {author} {\bibfnamefont {P.}~\bibnamefont
  {Strasberg}}, \bibinfo {author} {\bibfnamefont {G.}~\bibnamefont {Schaller}},
  \bibinfo {author} {\bibfnamefont {T.}~\bibnamefont {Brandes}}, \ and\
  \bibinfo {author} {\bibfnamefont {M.}~\bibnamefont {Esposito}},\ }\href
  {\doibase 10.1103/PhysRevX.7.021003} {\bibfield  {journal} {\bibinfo
  {journal} {Phys. Rev. X}\ }\textbf {\bibinfo {volume} {7}},\ \bibinfo {pages}
  {021003} (\bibinfo {year} {2017})}\BibitemShut {NoStop}%
\bibitem [{\citenamefont {Hanson}\ \emph {et~al.}(2018)\citenamefont {Hanson},
  \citenamefont {Joye}, \citenamefont {Pautrat},\ and\ \citenamefont
  {Raqu{\'e}pas}}]{Hanson2018}%
  \BibitemOpen
  \bibfield  {author} {\bibinfo {author} {\bibfnamefont {E.~P.}\ \bibnamefont
  {Hanson}}, \bibinfo {author} {\bibfnamefont {A.}~\bibnamefont {Joye}},
  \bibinfo {author} {\bibfnamefont {Y.}~\bibnamefont {Pautrat}}, \ and\
  \bibinfo {author} {\bibfnamefont {R.}~\bibnamefont {Raqu{\'e}pas}},\ }\href
  {\doibase 10.1007/s00023-018-0679-1} {\bibfield  {journal} {\bibinfo
  {journal} {Ann.~Henri Poincar{\'e}}\ }\textbf {\bibinfo {volume} {19}},\
  \bibinfo {pages} {1939} (\bibinfo {year} {2018})}\BibitemShut {NoStop}%
\bibitem [{\citenamefont {Grimmer}\ \emph
  {et~al.}(2018{\natexlab{a}})\citenamefont {Grimmer}, \citenamefont {Brown},
  \citenamefont {Kempf}, \citenamefont {Mann},\ and\ \citenamefont
  {Mart{\'{i}}n-Mart{\'{i}}nez}}]{Grimmer2018}%
  \BibitemOpen
  \bibfield  {author} {\bibinfo {author} {\bibfnamefont {D.}~\bibnamefont
  {Grimmer}}, \bibinfo {author} {\bibfnamefont {E.}~\bibnamefont {Brown}},
  \bibinfo {author} {\bibfnamefont {A.}~\bibnamefont {Kempf}}, \bibinfo
  {author} {\bibfnamefont {R.~B.}\ \bibnamefont {Mann}}, \ and\ \bibinfo
  {author} {\bibfnamefont {E.}~\bibnamefont {Mart{\'{i}}n-Mart{\'{i}}nez}},\
  }\href {\doibase 10.1103/PhysRevA.97.052120} {\bibfield  {journal} {\bibinfo
  {journal} {Phys. Rev. A}\ }\textbf {\bibinfo {volume} {97}},\ \bibinfo
  {pages} {052120} (\bibinfo {year} {2018}{\natexlab{a}})}\BibitemShut
  {NoStop}%
\bibitem [{\citenamefont {Grimmer}\ \emph
  {et~al.}(2018{\natexlab{b}})\citenamefont {Grimmer}, \citenamefont {Mann},\
  and\ \citenamefont {Martin-Martinez}}]{Grimmer2018b}%
  \BibitemOpen
  \bibfield  {author} {\bibinfo {author} {\bibfnamefont {D.}~\bibnamefont
  {Grimmer}}, \bibinfo {author} {\bibfnamefont {R.~B.}\ \bibnamefont {Mann}}, \
  and\ \bibinfo {author} {\bibfnamefont {E.}~\bibnamefont {Martin-Martinez}},\
  }\href {http://arxiv.org/abs/1805.11118} {\  (\bibinfo {year}
  {2018}{\natexlab{b}})},\ \Eprint {http://arxiv.org/abs/1805.11118}
  {arXiv:1805.11118} \BibitemShut {NoStop}%
\bibitem [{\citenamefont {{De Chiara}}\ \emph {et~al.}(2018)\citenamefont {{De
  Chiara}}, \citenamefont {Landi}, \citenamefont {Hewgill}, \citenamefont
  {Reid}, \citenamefont {Ferraro}, \citenamefont {Roncaglia},\ and\
  \citenamefont {Antezza}}]{deChiara2018}%
  \BibitemOpen
  \bibfield  {author} {\bibinfo {author} {\bibfnamefont {G.}~\bibnamefont {{De
  Chiara}}}, \bibinfo {author} {\bibfnamefont {G.}~\bibnamefont {Landi}},
  \bibinfo {author} {\bibfnamefont {A.}~\bibnamefont {Hewgill}}, \bibinfo
  {author} {\bibfnamefont {B.}~\bibnamefont {Reid}}, \bibinfo {author}
  {\bibfnamefont {A.}~\bibnamefont {Ferraro}}, \bibinfo {author} {\bibfnamefont
  {A.~J.}\ \bibnamefont {Roncaglia}}, \ and\ \bibinfo {author} {\bibfnamefont
  {M.}~\bibnamefont {Antezza}},\ }\href {\doibase 10.1088/1367-2630/aaecee}
  {\bibfield  {journal} {\bibinfo  {journal} {New J. Phys.}\ }\textbf {\bibinfo
  {volume} {20}},\ \bibinfo {pages} {113024} (\bibinfo {year}
  {2018})}\BibitemShut {NoStop}%
\bibitem [{\citenamefont {Barra}(2015)}]{barra2015}%
  \BibitemOpen
  \bibfield  {author} {\bibinfo {author} {\bibfnamefont {F.}~\bibnamefont
  {Barra}},\ }\href {\doibase 10.1038/srep14873} {\bibfield  {journal}
  {\bibinfo  {journal} {Sci.~Rep.}\ }\textbf {\bibinfo {volume} {5}},\ \bibinfo
  {pages} {14873} (\bibinfo {year} {2015})}\BibitemShut {NoStop}%
\bibitem [{\citenamefont {Hardal}\ and\ \citenamefont
  {M{\"u}stecapl{\i}o{\u{g}}lu}(2015)}]{hardal2015superradiant}%
  \BibitemOpen
  \bibfield  {author} {\bibinfo {author} {\bibfnamefont {A.~{\"U}.}\
  \bibnamefont {Hardal}}\ and\ \bibinfo {author} {\bibfnamefont {{\"O}.~E.}\
  \bibnamefont {M{\"u}stecapl{\i}o{\u{g}}lu}},\ }\href {\doibase
  10.1038/srep12953} {\bibfield  {journal} {\bibinfo  {journal} {Sci.~Rep.}\
  }\textbf {\bibinfo {volume} {5}},\ \bibinfo {pages} {12953} (\bibinfo {year}
  {2015})}\BibitemShut {NoStop}%
\bibitem [{\citenamefont {Manzano}\ \emph {et~al.}(2017)\citenamefont
  {Manzano}, \citenamefont {Silva},\ and\ \citenamefont
  {Parrondo}}]{manzano2017}%
  \BibitemOpen
  \bibfield  {author} {\bibinfo {author} {\bibfnamefont {G.}~\bibnamefont
  {Manzano}}, \bibinfo {author} {\bibfnamefont {R.}~\bibnamefont {Silva}}, \
  and\ \bibinfo {author} {\bibfnamefont {J.~M.}\ \bibnamefont {Parrondo}},\
  }\href {http://arxiv.org/abs/1709.00231} {\bibfield  {journal} {\bibinfo
  {journal} {arXiv:1709.00231}\ } (\bibinfo {year} {2017})}\BibitemShut
  {NoStop}%
\bibitem [{\citenamefont {Barra}\ and\ \citenamefont
  {Lled{\'o}}(2018)}]{Barra2018}%
  \BibitemOpen
  \bibfield  {author} {\bibinfo {author} {\bibfnamefont {F.}~\bibnamefont
  {Barra}}\ and\ \bibinfo {author} {\bibfnamefont {C.}~\bibnamefont
  {Lled{\'o}}},\ }\href {\doibase 10.1140/epjst/e2018-00084-x} {\bibfield
  {journal} {\bibinfo  {journal} {Eur.~Phys.~J. Special Topics}\ }\textbf
  {\bibinfo {volume} {227}},\ \bibinfo {pages} {231} (\bibinfo {year}
  {2018})}\BibitemShut {NoStop}%
\bibitem [{\citenamefont {Pezzutto}\ \emph {et~al.}(2018)\citenamefont
  {Pezzutto}, \citenamefont {Paternostro},\ and\ \citenamefont
  {Omar}}]{Pezzutto2018}%
  \BibitemOpen
  \bibfield  {author} {\bibinfo {author} {\bibfnamefont {M.}~\bibnamefont
  {Pezzutto}}, \bibinfo {author} {\bibfnamefont {M.}~\bibnamefont
  {Paternostro}}, \ and\ \bibinfo {author} {\bibfnamefont {Y.}~\bibnamefont
  {Omar}},\ }\href {http://arxiv.org/abs/1806.10075} {\bibfield  {journal}
  {\bibinfo  {journal} {arXiv:1806.10075}\ } (\bibinfo {year}
  {2018})}\BibitemShut {NoStop}%
\bibitem [{\citenamefont {Gallis}\ and\ \citenamefont
  {Fleming}(1990)}]{Gallis1990}%
  \BibitemOpen
  \bibfield  {author} {\bibinfo {author} {\bibfnamefont {M.~M.}\ \bibnamefont
  {Gallis}}\ and\ \bibinfo {author} {\bibfnamefont {G.~G.}\ \bibnamefont
  {Fleming}},\ }\href {\doibase 10.1103/PhysRevA.42.38} {\bibfield  {journal}
  {\bibinfo  {journal} {Phys. Rev. A}\ }\textbf {\bibinfo {volume} {42}},\
  \bibinfo {pages} {38} (\bibinfo {year} {1990})}\BibitemShut {NoStop}%
\bibitem [{\citenamefont {Diosi}(1995)}]{Diosi1995}%
  \BibitemOpen
  \bibfield  {author} {\bibinfo {author} {\bibfnamefont {L.}~\bibnamefont
  {Diosi}},\ }\href {\doibase 10.1209/0295-5075/30/2/001} {\bibfield  {journal}
  {\bibinfo  {journal} {Europhys.~Lett.}\ }\textbf {\bibinfo {volume} {30}},\
  \bibinfo {pages} {63} (\bibinfo {year} {1995})}\BibitemShut {NoStop}%
\bibitem [{\citenamefont {Hornberger}\ and\ \citenamefont
  {Sipe}(2003)}]{Hornberger2003c}%
  \BibitemOpen
  \bibfield  {author} {\bibinfo {author} {\bibfnamefont {K.}~\bibnamefont
  {Hornberger}}\ and\ \bibinfo {author} {\bibfnamefont {J.~E.}\ \bibnamefont
  {Sipe}},\ }\href {\doibase 10.1103/PhysRevA.68.012105} {\bibfield  {journal}
  {\bibinfo  {journal} {Phys. Rev. A}\ }\textbf {\bibinfo {volume} {68}},\
  \bibinfo {pages} {012105} (\bibinfo {year} {2003})}\BibitemShut {NoStop}%
\bibitem [{\citenamefont {Hornberger}(2006)}]{Hornberger2006}%
  \BibitemOpen
  \bibfield  {author} {\bibinfo {author} {\bibfnamefont {K.}~\bibnamefont
  {Hornberger}},\ }\href {\doibase 10.1103/PhysRevLett.97.060601} {\bibfield
  {journal} {\bibinfo  {journal} {Phys. Rev. Lett.}\ }\textbf {\bibinfo
  {volume} {97}},\ \bibinfo {pages} {060601} (\bibinfo {year}
  {2006})}\BibitemShut {NoStop}%
\bibitem [{\citenamefont {Hornberger}(2007)}]{Hornberger2007}%
  \BibitemOpen
  \bibfield  {author} {\bibinfo {author} {\bibfnamefont {K.}~\bibnamefont
  {Hornberger}},\ }\href {\doibase 10.1209/0295-5075/77/50007} {\bibfield
  {journal} {\bibinfo  {journal} {Europhys. Lett.}\ }\textbf {\bibinfo {volume}
  {77}},\ \bibinfo {pages} {50007} (\bibinfo {year} {2007})}\BibitemShut
  {NoStop}%
\bibitem [{\citenamefont {Vacchini}\ and\ \citenamefont
  {Hornberger}(2009)}]{Vacchini2009}%
  \BibitemOpen
  \bibfield  {author} {\bibinfo {author} {\bibfnamefont {B.}~\bibnamefont
  {Vacchini}}\ and\ \bibinfo {author} {\bibfnamefont {K.}~\bibnamefont
  {Hornberger}},\ }\href {\doibase 10.1016/j.physrep.2009.06.001} {\bibfield
  {journal} {\bibinfo  {journal} {Phys. Rep.}\ }\textbf {\bibinfo {volume}
  {478}},\ \bibinfo {pages} {71} (\bibinfo {year} {2009})}\BibitemShut
  {NoStop}%
\bibitem [{\citenamefont {Brand\~ao}\ \emph {et~al.}(2013)\citenamefont
  {Brand\~ao}, \citenamefont {Horodecki}, \citenamefont {Oppenheim},
  \citenamefont {Renes},\ and\ \citenamefont {Spekkens}}]{Brandao2013}%
  \BibitemOpen
  \bibfield  {author} {\bibinfo {author} {\bibfnamefont {F.}~\bibnamefont
  {Brand\~ao}}, \bibinfo {author} {\bibfnamefont {M.}~\bibnamefont
  {Horodecki}}, \bibinfo {author} {\bibfnamefont {J.}~\bibnamefont
  {Oppenheim}}, \bibinfo {author} {\bibfnamefont {J.~M.}\ \bibnamefont
  {Renes}}, \ and\ \bibinfo {author} {\bibfnamefont {R.~W.}\ \bibnamefont
  {Spekkens}},\ }\href {\doibase 10.1103/PhysRevLett.111.250404} {\bibfield
  {journal} {\bibinfo  {journal} {Phys. Rev. Lett.}\ }\textbf {\bibinfo
  {volume} {111}},\ \bibinfo {pages} {250404} (\bibinfo {year}
  {2013})}\BibitemShut {NoStop}%
\bibitem [{\citenamefont {Brand{\~{a}}o}\ \emph {et~al.}(2015)\citenamefont
  {Brand{\~{a}}o}, \citenamefont {Horodecki}, \citenamefont {Ng}, \citenamefont
  {Oppenheim},\ and\ \citenamefont {Wehner}}]{Brandao2015}%
  \BibitemOpen
  \bibfield  {author} {\bibinfo {author} {\bibfnamefont {F.}~\bibnamefont
  {Brand{\~{a}}o}}, \bibinfo {author} {\bibfnamefont {M.}~\bibnamefont
  {Horodecki}}, \bibinfo {author} {\bibfnamefont {N.}~\bibnamefont {Ng}},
  \bibinfo {author} {\bibfnamefont {J.}~\bibnamefont {Oppenheim}}, \ and\
  \bibinfo {author} {\bibfnamefont {S.}~\bibnamefont {Wehner}},\ }\href
  {\doibase 10.1073/pnas.1411728112} {\bibfield  {journal} {\bibinfo  {journal}
  {Proc. Natl. Acad. Sci.}\ }\textbf {\bibinfo {volume} {112}},\ \bibinfo
  {pages} {3275} (\bibinfo {year} {2015})}\BibitemShut {NoStop}%
\bibitem [{\citenamefont {Gour}\ \emph {et~al.}(2015)\citenamefont {Gour},
  \citenamefont {M{\"u}ller}, \citenamefont {Narasimhachar}, \citenamefont
  {Spekkens},\ and\ \citenamefont {Halpern}}]{Gour2015}%
  \BibitemOpen
  \bibfield  {author} {\bibinfo {author} {\bibfnamefont {G.}~\bibnamefont
  {Gour}}, \bibinfo {author} {\bibfnamefont {M.~P.}\ \bibnamefont
  {M{\"u}ller}}, \bibinfo {author} {\bibfnamefont {V.}~\bibnamefont
  {Narasimhachar}}, \bibinfo {author} {\bibfnamefont {R.~W.}\ \bibnamefont
  {Spekkens}}, \ and\ \bibinfo {author} {\bibfnamefont {N.~Y.}\ \bibnamefont
  {Halpern}},\ }\href {\doibase https://doi.org/10.1016/j.physrep.2015.04.003}
  {\bibfield  {journal} {\bibinfo  {journal} {Phys.~Rep.}\ }\textbf {\bibinfo
  {volume} {583}},\ \bibinfo {pages} {1 } (\bibinfo {year} {2015})}\BibitemShut
  {NoStop}%
\bibitem [{\citenamefont {Goold}\ \emph {et~al.}(2016)\citenamefont {Goold},
  \citenamefont {Huber}, \citenamefont {Riera}, \citenamefont {del Rio},\ and\
  \citenamefont {Skrzypczyk}}]{goold2016review}%
  \BibitemOpen
  \bibfield  {author} {\bibinfo {author} {\bibfnamefont {J.}~\bibnamefont
  {Goold}}, \bibinfo {author} {\bibfnamefont {M.}~\bibnamefont {Huber}},
  \bibinfo {author} {\bibfnamefont {A.}~\bibnamefont {Riera}}, \bibinfo
  {author} {\bibfnamefont {L.}~\bibnamefont {del Rio}}, \ and\ \bibinfo
  {author} {\bibfnamefont {P.}~\bibnamefont {Skrzypczyk}},\ }\href
  {http://stacks.iop.org/1751-8121/49/i=14/a=143001} {\bibfield  {journal}
  {\bibinfo  {journal} {J.~Phys.~A}\ }\textbf {\bibinfo {volume} {49}},\
  \bibinfo {pages} {143001} (\bibinfo {year} {2016})}\BibitemShut {NoStop}%
\bibitem [{\citenamefont {Lostaglio}\ \emph {et~al.}(2018)\citenamefont
  {Lostaglio}, \citenamefont {Alhambra},\ and\ \citenamefont
  {Perry}}]{Lostaglio2018}%
  \BibitemOpen
  \bibfield  {author} {\bibinfo {author} {\bibfnamefont {M.}~\bibnamefont
  {Lostaglio}}, \bibinfo {author} {\bibfnamefont {{\'{A}}.~M.}\ \bibnamefont
  {Alhambra}}, \ and\ \bibinfo {author} {\bibfnamefont {C.}~\bibnamefont
  {Perry}},\ }\href {\doibase 10.22331/q-2018-02-08-52} {\bibfield  {journal}
  {\bibinfo  {journal} {{Quantum}}\ }\textbf {\bibinfo {volume} {2}},\ \bibinfo
  {pages} {52} (\bibinfo {year} {2018})}\BibitemShut {NoStop}%
\bibitem [{\citenamefont {Yoshida}(1990)}]{Yoshida1990}%
  \BibitemOpen
  \bibfield  {author} {\bibinfo {author} {\bibfnamefont {H.}~\bibnamefont
  {Yoshida}},\ }\href {\doibase 10.1016/0375-9601(90)90092-3} {\bibfield
  {journal} {\bibinfo  {journal} {Phys. Lett. A}\ }\textbf {\bibinfo {volume}
  {150}},\ \bibinfo {pages} {262} (\bibinfo {year} {1990})}\BibitemShut
  {NoStop}%
\bibitem [{\citenamefont {Lidar}\ \emph {et~al.}(2006)\citenamefont {Lidar},
  \citenamefont {Shabani},\ and\ \citenamefont {Alicki}}]{Lidar2006}%
  \BibitemOpen
  \bibfield  {author} {\bibinfo {author} {\bibfnamefont {D.}~\bibnamefont
  {Lidar}}, \bibinfo {author} {\bibfnamefont {A.}~\bibnamefont {Shabani}}, \
  and\ \bibinfo {author} {\bibfnamefont {R.}~\bibnamefont {Alicki}},\ }\href
  {\doibase 10.1016/J.CHEMPHYS.2005.06.038} {\bibfield  {journal} {\bibinfo
  {journal} {Chem. Phys.}\ }\textbf {\bibinfo {volume} {322}},\ \bibinfo
  {pages} {82} (\bibinfo {year} {2006})}\BibitemShut {NoStop}%
\bibitem [{\citenamefont {Grimmer}\ \emph {et~al.}(2017)\citenamefont
  {Grimmer}, \citenamefont {Mann},\ and\ \citenamefont
  {Mart{\'{i}}n-Mart{\'{i}}nez}}]{Grimmer2017a}%
  \BibitemOpen
  \bibfield  {author} {\bibinfo {author} {\bibfnamefont {D.}~\bibnamefont
  {Grimmer}}, \bibinfo {author} {\bibfnamefont {R.~B.}\ \bibnamefont {Mann}}, \
  and\ \bibinfo {author} {\bibfnamefont {E.}~\bibnamefont
  {Mart{\'{i}}n-Mart{\'{i}}nez}},\ }\href {\doibase 10.1103/PhysRevA.95.042114}
  {\bibfield  {journal} {\bibinfo  {journal} {Phys. Rev. A}\ }\textbf {\bibinfo
  {volume} {95}},\ \bibinfo {pages} {042114} (\bibinfo {year}
  {2017})}\BibitemShut {NoStop}%
\bibitem [{\citenamefont {Breuer}\ and\ \citenamefont
  {Petruccione}(2002)}]{Breuer2002}%
  \BibitemOpen
  \bibfield  {author} {\bibinfo {author} {\bibfnamefont {H.-P.}\ \bibnamefont
  {Breuer}}\ and\ \bibinfo {author} {\bibfnamefont {F.}~\bibnamefont
  {Petruccione}},\ }\href
  {http://books.google.com/books?id=0Yx5VzaMYm8C{\&}pgis=1} {\emph {\bibinfo
  {title} {{The Theory of Open Quantum Systems}}}}\ (\bibinfo  {publisher}
  {Oxford University Press},\ \bibinfo {year} {2002})\BibitemShut {NoStop}%
\bibitem [{\citenamefont {Doll}\ \emph {et~al.}(2008)\citenamefont {Doll},
  \citenamefont {Zueco}, \citenamefont {Wubs}, \citenamefont {Kohler},\ and\
  \citenamefont {H\"{a}nggi}}]{Doll2008}%
  \BibitemOpen
  \bibfield  {author} {\bibinfo {author} {\bibfnamefont {R.}~\bibnamefont
  {Doll}}, \bibinfo {author} {\bibfnamefont {D.}~\bibnamefont {Zueco}},
  \bibinfo {author} {\bibfnamefont {M.}~\bibnamefont {Wubs}}, \bibinfo {author}
  {\bibfnamefont {S.}~\bibnamefont {Kohler}}, \ and\ \bibinfo {author}
  {\bibfnamefont {P.}~\bibnamefont {H\"{a}nggi}},\ }\href {\doibase
  https://doi.org/10.1016/j.chemphys.2007.09.003} {\bibfield  {journal}
  {\bibinfo  {journal} {Chem.~Phys.}\ }\textbf {\bibinfo {volume} {347}},\
  \bibinfo {pages} {243 } (\bibinfo {year} {2008})}\BibitemShut {NoStop}%
\bibitem [{\citenamefont {Jacobs}\ and\ \citenamefont
  {Steck}(2006)}]{jacobs2006}%
  \BibitemOpen
  \bibfield  {author} {\bibinfo {author} {\bibfnamefont {K.}~\bibnamefont
  {Jacobs}}\ and\ \bibinfo {author} {\bibfnamefont {D.~A.}\ \bibnamefont
  {Steck}},\ }\href {\doibase 10.1080/00107510601101934} {\bibfield  {journal}
  {\bibinfo  {journal} {Contemp.~Phys.}\ }\textbf {\bibinfo {volume} {47}},\
  \bibinfo {pages} {279} (\bibinfo {year} {2006})}\BibitemShut {NoStop}%
\bibitem [{\citenamefont {Wiseman}\ and\ \citenamefont
  {Milburn}(2009)}]{wiseman2009}%
  \BibitemOpen
  \bibfield  {author} {\bibinfo {author} {\bibfnamefont {H.~M.}\ \bibnamefont
  {Wiseman}}\ and\ \bibinfo {author} {\bibfnamefont {G.~J.}\ \bibnamefont
  {Milburn}},\ }\href {\doibase 10.1017/CBO9780511813948} {\emph {\bibinfo
  {title} {Quantum Measurement and Control}}}\ (\bibinfo  {publisher}
  {Cambridge University Press},\ \bibinfo {year} {2009})\BibitemShut {NoStop}%
\bibitem [{\citenamefont {Elouard}\ \emph
  {et~al.}(2017{\natexlab{a}})\citenamefont {Elouard}, \citenamefont
  {Herrera-Mart{\'\i}}, \citenamefont {Clusel},\ and\ \citenamefont
  {Auff{\`e}ves}}]{elouard2017}%
  \BibitemOpen
  \bibfield  {author} {\bibinfo {author} {\bibfnamefont {C.}~\bibnamefont
  {Elouard}}, \bibinfo {author} {\bibfnamefont {D.~A.}\ \bibnamefont
  {Herrera-Mart{\'\i}}}, \bibinfo {author} {\bibfnamefont {M.}~\bibnamefont
  {Clusel}}, \ and\ \bibinfo {author} {\bibfnamefont {A.}~\bibnamefont
  {Auff{\`e}ves}},\ }\href {\doibase 10.1038/s41534-017-0008-4} {\bibfield
  {journal} {\bibinfo  {journal} {npj Quantum Inf.}\ }\textbf {\bibinfo
  {volume} {3}},\ \bibinfo {pages} {9} (\bibinfo {year}
  {2017}{\natexlab{a}})}\BibitemShut {NoStop}%
\bibitem [{\citenamefont {Elouard}\ \emph
  {et~al.}(2017{\natexlab{b}})\citenamefont {Elouard}, \citenamefont
  {Herrera-Mart\'{\i}}, \citenamefont {Huard},\ and\ \citenamefont
  {Auff\`eves}}]{elouard2017maxwell}%
  \BibitemOpen
  \bibfield  {author} {\bibinfo {author} {\bibfnamefont {C.}~\bibnamefont
  {Elouard}}, \bibinfo {author} {\bibfnamefont {D.}~\bibnamefont
  {Herrera-Mart\'{\i}}}, \bibinfo {author} {\bibfnamefont {B.}~\bibnamefont
  {Huard}}, \ and\ \bibinfo {author} {\bibfnamefont {A.}~\bibnamefont
  {Auff\`eves}},\ }\href {\doibase 10.1103/PhysRevLett.118.260603} {\bibfield
  {journal} {\bibinfo  {journal} {Phys. Rev. Lett.}\ }\textbf {\bibinfo
  {volume} {118}},\ \bibinfo {pages} {260603} (\bibinfo {year}
  {2017}{\natexlab{b}})}\BibitemShut {NoStop}%
\bibitem [{\citenamefont {Yi}\ \emph {et~al.}(2017)\citenamefont {Yi},
  \citenamefont {Talkner},\ and\ \citenamefont {Kim}}]{yi2017}%
  \BibitemOpen
  \bibfield  {author} {\bibinfo {author} {\bibfnamefont {J.}~\bibnamefont
  {Yi}}, \bibinfo {author} {\bibfnamefont {P.}~\bibnamefont {Talkner}}, \ and\
  \bibinfo {author} {\bibfnamefont {Y.~W.}\ \bibnamefont {Kim}},\ }\href
  {\doibase 10.1103/PhysRevE.96.022108} {\bibfield  {journal} {\bibinfo
  {journal} {Phys. Rev. E}\ }\textbf {\bibinfo {volume} {96}},\ \bibinfo
  {pages} {022108} (\bibinfo {year} {2017})}\BibitemShut {NoStop}%
\bibitem [{\citenamefont {Elouard}\ and\ \citenamefont
  {Jordan}(2018)}]{elouard2018}%
  \BibitemOpen
  \bibfield  {author} {\bibinfo {author} {\bibfnamefont {C.}~\bibnamefont
  {Elouard}}\ and\ \bibinfo {author} {\bibfnamefont {A.~N.}\ \bibnamefont
  {Jordan}},\ }\href {\doibase 10.1103/PhysRevLett.120.260601} {\bibfield
  {journal} {\bibinfo  {journal} {Phys. Rev. Lett.}\ }\textbf {\bibinfo
  {volume} {120}},\ \bibinfo {pages} {260601} (\bibinfo {year}
  {2018})}\BibitemShut {NoStop}%
\bibitem [{\citenamefont {Buffoni}\ \emph {et~al.}(2018)\citenamefont
  {Buffoni}, \citenamefont {Solfanelli}, \citenamefont {Verrucchi},
  \citenamefont {Cuccoli},\ and\ \citenamefont {Campisi}}]{buffoni2018}%
  \BibitemOpen
  \bibfield  {author} {\bibinfo {author} {\bibfnamefont {L.}~\bibnamefont
  {Buffoni}}, \bibinfo {author} {\bibfnamefont {A.}~\bibnamefont {Solfanelli}},
  \bibinfo {author} {\bibfnamefont {P.}~\bibnamefont {Verrucchi}}, \bibinfo
  {author} {\bibfnamefont {A.}~\bibnamefont {Cuccoli}}, \ and\ \bibinfo
  {author} {\bibfnamefont {M.}~\bibnamefont {Campisi}},\ }\href
  {https://arxiv.org/abs/1806.07814} {\bibfield  {journal} {\bibinfo  {journal}
  {arXiv:1806.07814}\ } (\bibinfo {year} {2018})}\BibitemShut {NoStop}%
\bibitem [{\citenamefont {Allahverdyan}\ \emph {et~al.}(2004)\citenamefont
  {Allahverdyan}, \citenamefont {Balian},\ and\ \citenamefont
  {Nieuwenhuizen}}]{allah2004work}%
  \BibitemOpen
  \bibfield  {author} {\bibinfo {author} {\bibfnamefont {A.~E.}\ \bibnamefont
  {Allahverdyan}}, \bibinfo {author} {\bibfnamefont {R.}~\bibnamefont
  {Balian}}, \ and\ \bibinfo {author} {\bibfnamefont {T.~M.}\ \bibnamefont
  {Nieuwenhuizen}},\ }\href {http://stacks.iop.org/0295-5075/67/i=4/a=565}
  {\bibfield  {journal} {\bibinfo  {journal} {Europhys. Lett.}\ }\textbf
  {\bibinfo {volume} {67}},\ \bibinfo {pages} {565} (\bibinfo {year}
  {2004})}\BibitemShut {NoStop}%
\bibitem [{\citenamefont {Skrzypczyk}\ \emph {et~al.}(2011)\citenamefont
  {Skrzypczyk}, \citenamefont {Brunner}, \citenamefont {Linden},\ and\
  \citenamefont {Popescu}}]{skrzypczyk2011}%
  \BibitemOpen
  \bibfield  {author} {\bibinfo {author} {\bibfnamefont {P.}~\bibnamefont
  {Skrzypczyk}}, \bibinfo {author} {\bibfnamefont {N.}~\bibnamefont {Brunner}},
  \bibinfo {author} {\bibfnamefont {N.}~\bibnamefont {Linden}}, \ and\ \bibinfo
  {author} {\bibfnamefont {S.}~\bibnamefont {Popescu}},\ }\href
  {http://stacks.iop.org/1751-8121/44/i=49/a=492002} {\bibfield  {journal}
  {\bibinfo  {journal} {J.~Phys.~A}\ }\textbf {\bibinfo {volume} {44}},\
  \bibinfo {pages} {492002} (\bibinfo {year} {2011})}\BibitemShut {NoStop}%
\bibitem [{\citenamefont {Brask}\ \emph {et~al.}(2015)\citenamefont {Brask},
  \citenamefont {Haack}, \citenamefont {Brunner},\ and\ \citenamefont
  {Huber}}]{brask2015entangle}%
  \BibitemOpen
  \bibfield  {author} {\bibinfo {author} {\bibfnamefont {J.~B.}\ \bibnamefont
  {Brask}}, \bibinfo {author} {\bibfnamefont {G.}~\bibnamefont {Haack}},
  \bibinfo {author} {\bibfnamefont {N.}~\bibnamefont {Brunner}}, \ and\
  \bibinfo {author} {\bibfnamefont {M.}~\bibnamefont {Huber}},\ }\href
  {http://stacks.iop.org/1367-2630/17/i=11/a=113029} {\bibfield  {journal}
  {\bibinfo  {journal} {New J. Phys.}\ }\textbf {\bibinfo {volume} {17}},\
  \bibinfo {pages} {113029} (\bibinfo {year} {2015})}\BibitemShut {NoStop}%
\bibitem [{\citenamefont {Rivas}\ \emph {et~al.}(2010)\citenamefont {Rivas},
  \citenamefont {Plato}, \citenamefont {Huelga},\ and\ \citenamefont
  {Plenio}}]{rivas2010}%
  \BibitemOpen
  \bibfield  {author} {\bibinfo {author} {\bibfnamefont {A.}~\bibnamefont
  {Rivas}}, \bibinfo {author} {\bibfnamefont {A.~D.~K.}\ \bibnamefont {Plato}},
  \bibinfo {author} {\bibfnamefont {S.~F.}\ \bibnamefont {Huelga}}, \ and\
  \bibinfo {author} {\bibfnamefont {M.~B.}\ \bibnamefont {Plenio}},\ }\href
  {\doibase 10.1088/1367-2630/12/11/113032} {\bibfield  {journal} {\bibinfo
  {journal} {New J. Phys.}\ }\textbf {\bibinfo {volume} {12}},\ \bibinfo
  {pages} {113032} (\bibinfo {year} {2010})}\BibitemShut {NoStop}%
\bibitem [{\citenamefont {Levy}\ and\ \citenamefont
  {Kosloff}(2014)}]{levy2014}%
  \BibitemOpen
  \bibfield  {author} {\bibinfo {author} {\bibfnamefont {A.}~\bibnamefont
  {Levy}}\ and\ \bibinfo {author} {\bibfnamefont {R.}~\bibnamefont {Kosloff}},\
  }\href {\doibase 10.1209/0295-5075/107/20004} {\bibfield  {journal} {\bibinfo
   {journal} {Europhys. Lett.}\ }\textbf {\bibinfo {volume} {107}},\ \bibinfo
  {pages} {20004} (\bibinfo {year} {2014})}\BibitemShut {NoStop}%
\bibitem [{\citenamefont {Hofer}\ \emph {et~al.}(2017)\citenamefont {Hofer},
  \citenamefont {Perarnau-Llobet}, \citenamefont {Miranda}, \citenamefont
  {Haack}, \citenamefont {Silva}, \citenamefont {Brask},\ and\ \citenamefont
  {Brunner}}]{hofer2017}%
  \BibitemOpen
  \bibfield  {author} {\bibinfo {author} {\bibfnamefont {P.~P.}\ \bibnamefont
  {Hofer}}, \bibinfo {author} {\bibfnamefont {M.}~\bibnamefont
  {Perarnau-Llobet}}, \bibinfo {author} {\bibfnamefont {L.~D.~M.}\ \bibnamefont
  {Miranda}}, \bibinfo {author} {\bibfnamefont {G.}~\bibnamefont {Haack}},
  \bibinfo {author} {\bibfnamefont {R.}~\bibnamefont {Silva}}, \bibinfo
  {author} {\bibfnamefont {J.~B.}\ \bibnamefont {Brask}}, \ and\ \bibinfo
  {author} {\bibfnamefont {N.}~\bibnamefont {Brunner}},\ }\href {\doibase
  10.1088/1367-2630/aa964f} {\bibfield  {journal} {\bibinfo  {journal} {New J.
  Phys.}\ }\textbf {\bibinfo {volume} {19}},\ \bibinfo {pages} {123037}
  (\bibinfo {year} {2017})}\BibitemShut {NoStop}%
\bibitem [{\citenamefont {Gonz{\'a}lez}\ \emph {et~al.}(2017)\citenamefont
  {Gonz{\'a}lez}, \citenamefont {Correa}, \citenamefont {Nocerino},
  \citenamefont {Palao}, \citenamefont {Alonso},\ and\ \citenamefont
  {Adesso}}]{gonzalez2017}%
  \BibitemOpen
  \bibfield  {author} {\bibinfo {author} {\bibfnamefont {J.~O.}\ \bibnamefont
  {Gonz{\'a}lez}}, \bibinfo {author} {\bibfnamefont {L.~A.}\ \bibnamefont
  {Correa}}, \bibinfo {author} {\bibfnamefont {G.}~\bibnamefont {Nocerino}},
  \bibinfo {author} {\bibfnamefont {J.~P.}\ \bibnamefont {Palao}}, \bibinfo
  {author} {\bibfnamefont {D.}~\bibnamefont {Alonso}}, \ and\ \bibinfo {author}
  {\bibfnamefont {G.}~\bibnamefont {Adesso}},\ }\href {\doibase
  10.1142/S1230161217400108} {\bibfield  {journal} {\bibinfo  {journal} {Open
  Syst.~Inf.~Dyn.}\ }\textbf {\bibinfo {volume} {24}},\ \bibinfo {pages}
  {1740010} (\bibinfo {year} {2017})}\BibitemShut {NoStop}%
\bibitem [{\citenamefont {Cresser}\ and\ \citenamefont
  {Facer}(2017)}]{cresser2017coarse}%
  \BibitemOpen
  \bibfield  {author} {\bibinfo {author} {\bibfnamefont {J.}~\bibnamefont
  {Cresser}}\ and\ \bibinfo {author} {\bibfnamefont {C.}~\bibnamefont
  {Facer}},\ }\href {https://arxiv.org/abs/1710.09939} {\bibfield  {journal}
  {\bibinfo  {journal} {arXiv:1710.09939}\ } (\bibinfo {year}
  {2017})}\BibitemShut {NoStop}%
\bibitem [{\citenamefont {Seah}\ \emph {et~al.}(2018)\citenamefont {Seah},
  \citenamefont {Nimmrichter},\ and\ \citenamefont
  {Scarani}}]{stella2018fridge}%
  \BibitemOpen
  \bibfield  {author} {\bibinfo {author} {\bibfnamefont {S.}~\bibnamefont
  {Seah}}, \bibinfo {author} {\bibfnamefont {S.}~\bibnamefont {Nimmrichter}}, \
  and\ \bibinfo {author} {\bibfnamefont {V.}~\bibnamefont {Scarani}},\ }\href
  {\doibase 10.1103/PhysRevE.98.012131} {\bibfield  {journal} {\bibinfo
  {journal} {Phys. Rev. E}\ }\textbf {\bibinfo {volume} {98}},\ \bibinfo
  {pages} {012131} (\bibinfo {year} {2018})}\BibitemShut {NoStop}%
\bibitem [{\citenamefont {Cresser}\ \emph {et~al.}(2006)\citenamefont
  {Cresser}, \citenamefont {Barnett}, \citenamefont {Jeffers},\ and\
  \citenamefont {Pegg}}]{cresser2006}%
  \BibitemOpen
  \bibfield  {author} {\bibinfo {author} {\bibfnamefont {J.~D.}\ \bibnamefont
  {Cresser}}, \bibinfo {author} {\bibfnamefont {S.~M.}\ \bibnamefont
  {Barnett}}, \bibinfo {author} {\bibfnamefont {J.}~\bibnamefont {Jeffers}}, \
  and\ \bibinfo {author} {\bibfnamefont {D.~T.}\ \bibnamefont {Pegg}},\ }\href
  {\doibase https://doi.org/10.1016/j.optcom.2006.02.061} {\bibfield  {journal}
  {\bibinfo  {journal} {Opt.~Commun.}\ }\textbf {\bibinfo {volume} {264}},\
  \bibinfo {pages} {352 } (\bibinfo {year} {2006})}\BibitemShut {NoStop}%
\bibitem [{\citenamefont {Strasberg}(2018)}]{strasberg2018}%
  \BibitemOpen
  \bibfield  {author} {\bibinfo {author} {\bibfnamefont {P.}~\bibnamefont
  {Strasberg}},\ }\href {https://arxiv.org/abs/1810.00698} {\bibfield
  {journal} {\bibinfo  {journal} {arXiv:1810.00698}\ } (\bibinfo {year}
  {2018})}\BibitemShut {NoStop}%
\bibitem [{\citenamefont {Lidar}\ \emph {et~al.}(2001)\citenamefont {Lidar},
  \citenamefont {Bihary},\ and\ \citenamefont {Whaley}}]{lidar2001}%
  \BibitemOpen
  \bibfield  {author} {\bibinfo {author} {\bibfnamefont {D.~A.}\ \bibnamefont
  {Lidar}}, \bibinfo {author} {\bibfnamefont {Z.}~\bibnamefont {Bihary}}, \
  and\ \bibinfo {author} {\bibfnamefont {K.}~\bibnamefont {Whaley}},\ }\href
  {\doibase https://doi.org/10.1016/S0301-0104(01)00330-5} {\bibfield
  {journal} {\bibinfo  {journal} {Chem.~Phys.}\ }\textbf {\bibinfo {volume}
  {268}},\ \bibinfo {pages} {35 } (\bibinfo {year} {2001})}\BibitemShut
  {NoStop}%
\end{thebibliography}
%merlin.mbs apsrev4-1.bst 2010-07-25 4.21a (PWD, AO, DPC) hacked
%Control: key (0)
%Control: author (8) initials jnrlst
%Control: editor formatted (1) identically to author
%Control: production of article title (-1) disabled
%Control: page (0) single
%Control: year (1) truncated
%Control: production of eprint (0) enabled
%

\end{document}